\documentclass[12pt,preprint]{aastex}         % this makes it ApJ pre-print format for submission

\newcommand{\beq}{\begin{equation}}
\newcommand{\eeq}{\end{equation}}
\newcommand{\lan}{\langle}
\newcommand{\ran}{\rangle} 
\begin{document}
\title{The Polatron:  A Millimeter-Wave Cosmic Microwave Background Polarimeter for the OVRO 5.5 m Telescope} 
\author{B.J. Philhour\altaffilmark{1,2}, B.G. Keating\altaffilmark{1}, P.A.R. Ade\altaffilmark{3}, R.S. Bhatia\altaffilmark{1}, J.J. Bock\altaffilmark{1,4}, S.E. Church\altaffilmark{2}, J. Glenn\altaffilmark{5}, J.R. Hinderks\altaffilmark{2}, V.V. Hristov\altaffilmark{1}, W.C. Jones\altaffilmark{1}, M. Kamionkowski\altaffilmark{1}, D.E. Kumar\altaffilmark{1}, A.E. Lange\altaffilmark{1}, J.R. Leong\altaffilmark{1}, D.P. Marrone\altaffilmark{6}, B.S. Mason\altaffilmark{1}, P.V. Mason\altaffilmark{1,4}, M.M. Shuman\altaffilmark{1}, G.I. Sirbi\altaffilmark{1}}
\altaffiltext{1}{California Institute of Technology, Observational Cosmology, M.S. 59--33, Pasadena, CA 91125 \textit{bjp@astro.caltech.edu}}
\altaffiltext{2}{Stanford University, Varian Building, Stanford, CA 94305}
\altaffiltext{3}{Queen Mary and Westfield College, Astrophysics Laboratory, Department of Physics, London E1 4NS, UK}
\altaffiltext{4}{Jet Propulsion Laboratory, 4800 Oak Grove Dr., Pasadena, CA  91109}
\altaffiltext{5}{University of Colorado, Center for Astrophysics and Space Astronomy, Boulder, CO 80309}
\altaffiltext{6}{University of Minnesota, School of Physics and Astronomy, Minneapolis, MN 55455}
\shorttitle{The Polatron:  A mm-Wave CMB Polarimeter}
\shortauthors{Philhour, B.J., et al.}
\begin{abstract}
We describe the development of a bolometric receiver designed to measure the arcminute-scale polarization of the cosmic microwave background (CMB).  The Polatron will be mounted at the Cassegrain focus of the 5.5 m telescope at the Owens Valley Radio Observatory (OVRO).  The receiver will measure both the \textit{Q} and \textit{U} Stokes parameters over a 20\% pass-band centered near 100 GHz, with the input polarization signal modulated at $\sim 0.6$ Hz by a rotating, birefringent, quartz half-wave plate. In six months of observation we plan to observe $\sim 400$ 2.5 arcminute pixels in a ring about the North Celestial Pole to a precision of $\sim 6~\mu K$/pixel in each of \textit{Q} and \textit{U}, adequate to unambiguously detect CMB polarization at levels predicted by current models. 
\end{abstract} 

\section{Introduction}
The detailed structure of the angular power spectrum of the cosmic microwave background is currently being used to constrain a variety of cosmological parameters (see, e.g., \citet{lange00}).  Thomson scattering of local quadrupole distributions of this radiation at the surface of last scattering should generate linearly polarized signal \citep{chandra60}.  Consequently, the temperature anisotropies that are induced by density fluctuations and gravity waves in the early universe should themselves induce measurable polarization of the CMB (see, e.g., \citet{jaffe00}).  

Measurement of the pattern and amplitude of the CMB polarization field should ultimately allow tests of a number of cosmological theories.  First, it will enable a check on the consistency of the theoretical model that describes CMB $\Delta T$ fluctuations as arising from density fluctuations in the early universe; the cosmological models that predict the temperature angular power spectra predict corresponding polarization angular power spectra.  Second, information gleaned from the polarization angular power spectrum will break degeneracies between cosmological parameters that arise in interpreting $\Delta T$ fluctuations.  For example, it is difficult to use $\Delta T$ information alone to distinguish between dilution of signal at large angular scales due to early reionization and an overall change in the normalization of the primordial power spectrum.  However, early reionization should produce a recognizable signal --- \textit{more} power at large angular scales --- in polarization (see \citet{keating98}).  Third, a curl component to the CMB polarization vector pattern would indicate the presence of the stochastic gravity-wave background predicted by inflation models \citep{kamionkowski97, seljak97, kamionkowski97b}.  Should they exist, primordial gravity waves would carry information from an inflationary era never before open to investigation.  

The current observational status of this developing field is analogous to that of temperature anisotropy measurements some ten to fifteen years ago, with even the most sensitive upper limits an order of magnitude above the theoretically expected signal.  This situation is set to change very soon with the ongoing deployment of a new generation of experiments specifically designed to measure CMB polarization.

\subsection{Polarization Angular Power Spectra}
Because of its vector nature, the formalism through which we compare measurement and theory is more complex for polarization than for intensity anisotropies.  We begin with a CMB photon traveling in the $\hat{z}$ direction with wavelength $\lambda$ and frequency $\nu$, with an electric field vector
\begin{eqnarray}
\overrightarrow{E}(x,y,z,t) & = & E_{x}(t)\sin\left[2\pi\left(\frac{z}{\lambda} - \nu t\right) + \phi_{x}(t)\right] \hat{x} \nonumber \\ 
& & + E_{y}(t)\sin\left[2\pi\left(\frac{z}{\lambda} - \nu t\right) + \phi_{y}(t)\right] \hat{y}.
\end{eqnarray}
The degree to which $\phi_{x}(t)$ and $\phi_{y}(t)$ are correlated is the degree to which the light is polarized.  Time-averaging over the electric field oscillations, we can describe the polarization with the Stokes parameters
\begin{eqnarray}
I & \equiv & \langle E^{2}_{x} \rangle + \langle E^{2}_{y} \rangle, \\
Q & \equiv & \langle E^{2}_{x} \rangle - \langle E^{2}_{y} \rangle, \\
U & \equiv & \langle 2 E_{x} E_{y} \cos (\phi_{x} - \phi_{y}) \rangle, \\
V & \equiv & \langle 2 E_{x} E_{y} \sin (\phi_{x} - \phi_{y}) \rangle,
\end{eqnarray}
where $I$ refers to the intensity, $Q$ and $U$ describe the linear polarization and depend on the definition of the $(x,y)$ coordinate system, and $V$ describes the circular polarization of the photon.  Thomson scattering does not produce circular polarization.  Hence, the CMB is not expected to be circularly polarized, and we no longer consider $V$.  The Stokes parameters are converted from units of flux density to CMB brightness temperature through cross-calibration with an astrophysical source of known temperature and emissivity.

Maps for which the CMB temperature $T$ is known in each direction $\hat{n}$ are routinely analyzed via expansion of the map in terms of the spherical harmonics $Y_{lm}$:
\begin{equation}
\label{exp}
\frac{T(\hat{n})}{T_{cmb}} = 1 + \sum_{l=1}^{\infty} \sum_{m=-l}^{l} a^{T}_{lm} Y_{lm} (\hat{n}),
\end{equation}
where
\begin{equation}
a^{T}_{lm} = \frac{1}{T_{cmb}} \int d\hat{n} \,T(\hat{n}) Y^{*}_{lm}(\hat{n}),
\end{equation}
and $T_{cmb}$ is the average temperature of the CMB.  The auto-correlation function $C_{l}^{T}$, defined by
\begin{equation}
\langle a^{T\,*}_{lm}~a^{T}_{l'm'} \rangle = C_{l}^{T} \delta_{ll'} \delta_{mm'},
\end{equation}
is predicted by theory.  The multipole moments $a^{T}_{lm}$ are a realization of the underlying statistical theory that are specific to our universe and our unique viewpoint within it.  

\citet{kamionkowski97} have developed a formalism very similar to that described above for the analysis of polarization maps.  (A separate and equivalent formalism has been developed by \citet{zaldarriaga97}.)  They define a symmetric, trace-free tensor $P_{ab}(\hat{n})$ which describes the linear polarization observed in direction $\hat{n} = (\theta, \phi)$,
\begin{equation}
\label{pab}
P_{ab}(\hat{n}) = \frac{1}{2} \left(
\begin{array}{cc}
Q(\hat{n})              & -U(\hat{n})\sin\theta \\
-U(\hat{n})\sin\theta & -Q(\hat{n})\sin^{2}\theta 
\end{array}
\right).
\end{equation}
By analogy to (\ref{exp}), $P_{ab}(\hat{n})$ is expanded as a sum of appropriate orthonormal basis functions on the sphere,
\begin{equation}
\frac{P_{ab}(\hat{n})}{T_{cmb}} = \sum_{l=1}^{\infty} \sum_{m=-l}^{l} \left[ a^{G}_{lm} Y^{G}_{(lm)ab} (\hat{n})
 + a^{C}_{lm} Y^{C}_{(lm)ab} (\hat{n}) \right],
\end{equation}
where
\begin{eqnarray}
a^{G}_{lm} = \frac{1}{T_{cmb}} \int d\hat{n} \, P_{ab}(\hat{n})Y_{lm}^{G\,ab\,*}(\hat{n}), \\
a^{C}_{lm} = \frac{1}{T_{cmb}} \int d\hat{n} \, P_{ab}(\hat{n})Y_{lm}^{C\,ab\,*}(\hat{n}),
\end{eqnarray}
and the basis functions $Y^{G}_{(lm)ab}$ and $Y^{C}_{(lm)ab}$ are covariant second derivatives on the sphere of the usual spherical harmonics $Y_{lm}$, chosen because they are a complete, orthonormal basis set for symmetric, trace-free tensors.  The expansion of the polarization tensor is broken into two natural sets of basis functions, labeled $G$ and $C$, since a symmetric, trace-free $2\times2$ tensor such as $P_{ab}$ can be written as a sum of two tensors, one with ``electric'' or ``gradient'' parity $(-1)^{l}$ and one with ``magnetic'' or ``curl'' parity $(-1)^{l+1}$.

The multipole moments $a^{G}_{lm}$, $a^{C}_{lm}$, and the temperature multipole moments $a^{T}_{lm}$ should completely describe a map of the CMB in both temperature and polarization.  An expanded set of power spectra is needed to fully characterize the temperature and polarization state of the CMB:
\begin{eqnarray}
\langle a^{T\,*}_{lm}~a^{T}_{l'm'} \rangle & = & C_{l}^{T} \delta_{ll'} \delta_{mm'}, \\
\langle a^{T\,*}_{lm}~a^{G}_{l'm'} \rangle & = & C_{l}^{TG} \delta_{ll'} \delta_{mm'}, \\
\langle a^{G\,*}_{lm}~a^{G}_{l'm'} \rangle & = & C_{l}^{G} \delta_{ll'} \delta_{mm'}, \\
\langle a^{G\,*}_{lm}~a^{C}_{l'm'} \rangle & = & C_{l}^{GC} \delta_{ll'} \delta_{mm'}, \\
\langle a^{C\,*}_{lm}~a^{C}_{l'm'} \rangle & = & C_{l}^{C} \delta_{ll'} \delta_{mm'}, \\
\langle a^{C\,*}_{lm}~a^{T}_{l'm'} \rangle & = & C_{l}^{CT} \delta_{ll'} \delta_{mm'}.
\end{eqnarray}

If the universe prefers no direction for CMB polarization, $C_{l}^{GC}$ and $C_{l}^{CT}$ are zero.  The remaining $C_{l}$'s encode fundamental cosmological parameters.  Generation of temperature anisotropies through scalar processes such as gravitational collapse will produce only $G$-type polarization since there is no preferred handedness or curl direction for such processes.  Temperature anisotropies generated through tensor (and vector) processes such as the influence of primordial gravity waves will produce both $G$- and $C$-type polarization \citep{kamionkowski97, seljak97}.

The reader is directed to \citet{kamionkowski97} for an extension of this formalism which allows for the calculation of multipole moments and variances given measured polarization two-point correlation functions.  

\subsection{Measurements to Date}
\citet{hedman00} have placed the strongest limits on $C_{l}^{G}$ and $C_{l}^{C}$ with a ground-based correlation polarimeter, PIQUE.  To simplify the comparison of one experimental result to another, polarization measurements are often expressed in terms of ``flat band power'' temperatures, such that $l(l+1)C^{X}_{l} / 2\pi = T^{2}_{X}$ where $X$ refers to $G$- and $C$-type polarization.  In the multipole windows $\langle l_{G} \rangle = 211^{+294}_{-146}$ and $\langle l_{C} \rangle = 212^{+229}_{-135}$ they place 95\% confidence limits on the flat band power of the $G$- and $C$-type power spectra of 10 and 9 $\mu K$ respectively, and a limit on the $G$-type spectrum of 7 $\mu K$ if $C_{l}^{C}$ is assumed to be zero.  These results have been divided by $\sqrt{2}$ to account for the difference between the G \& C formalism for polarization power spectra used here and the E \& B formalism used by \citet{hedman00}.  This upper limit, along with predicted angular power spectra for a chosen cosmological model, is plotted in Figure 1.

The Saskatoon experiment \citep{netterfield95} put a 95\% confidence upper limit of 16 $\mu K$ on CMB polarization in the angular scale range corresponding to $50 \leq l \leq 100$.  At smaller angular scales, $1' \leq \theta \leq 3'$, \citet{partridge88} put a 95\% confidence upper limit of 100 $\mu K$ on polarization.  (These data have not been reanalyzed to provide meaningful band power limits for inclusion in Figure 1.) 

Two satellite-borne observatories intended to measure temperature anisotropies, but which also have sensitivity to polarization, are under construction.  NASA's Microwave Anisotropy Probe (MAP, see \url{map.fsgc.nasa.gov}), to be launched in 2001, and ESA's Planck Surveyor (see \url{astro.estec.esa.nl/SA-general/Projects/Planck}), to be launched in 2007, will create full-sky maps in polarization at many frequencies.  The High Frequency Instrument (HFI) on the Planck Surveyor should be sensitive enough to allow polarization angular power spectroscopy.  

In advance of these observatories, many projects are planned or already underway.  Experimental techniques vary greatly, with significantly different detector systems (incoherent, coherent correlation, and coherent Dicke-switched polarimeters), polarization analyzers (orthomode transducers, wire grids, polarization-sensitive bolometers, and wave plates), observing frequencies (between 15 and 400 GHz), telescopes (upward-looking horns, single on- and off-axis dishes, and interferometers), beam sizes (from $7^{\circ}$ to 2.5'), observing sites (balloon- and satellite-borne as well as ground-based), and scan strategies (switched, scanned, and drift-scanned; observations of rings, patches, or the entire sky).  For a review of some current polarization experiments, see \citet{staggs99}.

\section{The Polatron:  Experimental Overview}
The Polatron is a bolometric polarimeter for the 5.5 m telescope at the Owens Valley Radio Observatory (OVRO) designed to measure CMB polarization on angular scales where the signal is expected to be highest, in a spectral band where galactic foregrounds are expected to be lowest.  The radiometer shares many elements of design with the SuZIE and Boomerang experiments \citep{holzapfel97, deBernardis99} and ESA's Planck Surveyor.  A single entrance feed is coupled to an orthomode transducer (OMT), which separates the two states of linear polarization.  Each OMT output is spectrally filtered and then terminated in a silicon-nitride micromesh bolometer.  The bolometer signals are electronically differenced, producing a signal that is proportional to the difference in optical power in the two arms of the OMT.  A half-wave plate placed in front of the receiver continuously rotates the incoming plane of polarization, allowing us to alternate between measurements of $Q$ and $U$ from the sky at a modulation frequency we choose.  A synopsis of the experimental specifications can be found in Table 1.  A schematic of the polarimeter can be found in Figure 2.  

\subsection{Optics Design}
The design of the Polatron optics aims to minimize background loading and systematic polarization effects while efficiently coupling the detectors to the sky with a beam matched to angular scales of scientific interest.  The OVRO 5.5 m telescope is ideally suited for such observation.  From 1992 to 1997, this telescope was dedicated to measurements of primary and secondary CMB anisotropies as described in \citet{herbig95}, \citet{myers97}, \citet{leitch00}, and \citet{mason01}.  As part of this program, the telescope was modified to minimize the warm ground emission that couples into the receiver by scattering from the secondary mirror support legs, resulting in a measured ground spillover temperature of 9 K \citep{lawrence94}. 

The telescope has never before been used at wavelengths $< 1$ cm.  Although the surface is sufficiently smooth on wavelength scales, large-scale dish deformities might distort and/or dilute our beam.  The dish consists of sixteen panels, each of which is deformable and supported in nine places with bolts and calibrated brass shims.  When the telescope was assembled at its original Table Mountain site, the surface was measured and set to 0.20 mm \textit{rms} accuracy.  When the telescope was later disassembled and moved to the Owens Valley, the shim thicknesses were recorded and reproduced, but upon reassembly no new surface measurement was made.  In February, 1999, we measured the shape of the dish with an array of high-resolution capacitive sensors mounted on a rotating parabolic arm.  On $\gtrsim 10$ cm length scales, the measured surface \textit{rms} accuracy was $0.24 \pm 0.02$ mm or $\sim\lambda/13$ for $\lambda = 3$ mm.  Removal and addition of shims under the surface allowed us to reduce the \textit{rms} surface accuracy to $0.15 \pm 0.02$ mm or $\sim\lambda/20$.

Upon reflection off a metal surface, unpolarized light becomes partially polarized in a direction along the reflecting surface \citep{hecht82}.  For the type of off-axis telescopes favored for low-background CMB $\Delta T$ measurements, large polarization offsets arise which may depend on pointing direction and dish temperature. In order to minimize the number of reflections, particularly off-axis reflections, the Polatron receiver will be placed directly at the Cassegrain focus of the telescope.  

A single corrugated entrance feed illuminates the primary mirror with an edge taper of --20 db, reducing expected emission from ground spillover to $<5$ K.  Together with the $f/3.2$ optical design of the telescope, this requires use of a large entrance feed, 4.1 cm in aperture diameter and 16.3 cm in length.  In front of the entrance feed is a rotating, ambient temperature, birefringent, crystalline quartz half-wave plate 7.6 cm in diameter and 3.26 cm thick.  The wave plate generates a $\pi$ phase retardation between electric field vectors with wavelength $\sim 3$ mm incident on the fast and slow refraction axes of the quartz.  Consequently, the polarization pattern from the sky is rotated about the optic axis at four times the physical rotation rate of the wave plate.  This modulation technique is employed routinely in millimeter-wave polarimetry as it allows both Stokes parameters \textit{Q} and \textit{U} to be measured with a single horn \citep{glenn97, murray92}.

\subsubsection{Polarization Efficiency}
The \textit{polarization efficiency} of an optical element is the degree to which a 100\% polarized signal will remain polarized on passage through the element.  Two effects are expected to reduce the polarization efficiency of the wave plate.  First, observations are made over a broad spectral band, while the wave plate is fundamentally chromatic.  Second, the optics of the telescope are moderately fast, $f/3.2$.  Different parts of our beam travel different path lengths through the quartz, so a phase retardation other than $\pi$ will result for those parts of the beam.  In order to mitigate this second effect, the wave plate is placed in the near field of the feed horn, where the beam is most collimated.  One can calculate the extent to which an input polarized signal from the sky will be depolarized by these two effects.  In this system such a calculation predicts approximately 97\% polarization efficiency (see Appendix A).  This efficiency effect can be measured and corrected for with greater accuracy than our anticipated overall calibration uncertainty, and should not be confused with sources of systematic polarization discussed later.    

\subsection{Spectral Filtering}
Bolometers are sensitive to radiation over a broad range of frequencies.  We have chosen a 20\% FWHM spectral bandpass centered on 96 GHz.  The 88 GHz low-frequency edge of our band is set by a $> 3\lambda$ length of waveguide at the exit aperture of the feed horn.  It is difficult to match the corrugated throat of the entrance feed to such a small diameter, so the exit aperture of the feed is reduced to the choke diameter in quarter-wavelength steps.  The 106 GHz high-frequency edge of our band is set by a metal-mesh resonant grid filter \citep{Lee96}.  The filters are more efficient in free space than in waveguide, so we couple the outputs of the OMT to the grid filters via rectangular-to-circular waveguide transitions and $f/4$ conical feedhorns.  A second $f/4$ conical feedhorn for each channel reconcentrates the radiation into integrating cavities containing the bolometers.  High density polyethylene (HDPE) lenses at the large aperture of each $f/4$ feed collimate the radiation for passage through the filter at high incidence angle and also ensure that the two feeds couple efficiently \citep{Church96}.  The gap between the two $f/4$ feeds also allows us to cool the majority of the focal plane to 4 K while the bolometer back-end is cooled to 0.25 K; a similar thermal gap in waveguide would present optical alignment difficulties.  

The band-defining metal-mesh filters typically have a leak at twice their edge cut-off frequency, so an additional metal-mesh filter with an edge at 160 GHz is included.  A further low-pass alkali-halide filter coated with black polypropylene blocks radiation above 1650 GHz.  Both of these filters are cooled to 4.2 K.  Large diameter low-pass metal-mesh filters with edges at 240 and 300 GHz are placed in front of the feedhorn, at 77 K and 4.2 K, in order to limit the thermal load from the entrance window on the band-defining filters and the cryogenic stages.  The entrance window itself is a piece of HDPE with thickness tuned to prevent reflective loss.

None of the spectral filters is expected to produce systematic polarization of incoming signals.  However, the polarization properties of the band-defining filters are immaterial, since the two polarization components of the incoming radiation have already been separated by the OMT.  Systematic polarization induced by the two blocking filters and window in front of the feed horn will remain fixed with respect to the instrument while the polarization pattern from the sky rotates with the wave plate.  As discussed later, this effect is removed upon demodulation of our signal at four times the wave plate rotation frequency.

\subsection{Detectors}
The Polatron uses silicon-nitride micromesh (``spider-web'') bolometers placed in an AC-bridge \citep{wilbanks90}.  They are sine-wave biased at $200$ Hz.  The relative bias levels can be adjusted to trim out gain mismatch between the two channels.  The output signals are buffered by a matched, 120 K, low-noise JFET pair, then differenced and demodulated in ambient temperature, low-noise amplifier electronics, producing a signal proportional to the difference in optical power in the two polarizations:
\begin{eqnarray}
V_{DIFF} & = & S \times \eta_{opt} \eta_{pol} \left[ Q \cos(2\pi \times 4 f_{wp} ~t) \right. \nonumber \\
         & & + \left. U \sin(2\pi \times 4 f_{wp} ~t) \right]  +  V_{noise}.
\end{eqnarray}
Here, $S$ is the detector responsivity, $\eta_{opt}$ is the band-averaged optical (photon) efficiency of the instrument, $\eta_{pol}$ is the polarization efficiency, $Q$ and $U$ are fixed to the sky and convolved with our beam, $f_{wp}$ is the physical rotation rate of the wave plate, and $V_{noise}$ is the sum of detector noise and noise from the JFET amplifier.  An additional two bolometers with the same thermal and electrical properties and amplifier chain, but which are not exposed to light, are included in the system.  These ``dark'' bolometers are a useful diagnostic for a number of systematic effects that could impair our measurement.

The 0.3 K operating temperature of the bolometers is provided by a triple-stage $^{4}$He/$^{3}$He/$^{3}$He sorption refrigerator \citep{bhatia01}.  The $^{4}$He stage condensation point temperature is set by a 4.2 K cryocooler purchased from APD Cryogenics.  A study carried out by \citet{bhatia99} of the susceptibility of bolometer systems to cryocooler microphonics suggests that cryocoolers are a feasible alternative to liquid He dewars for remote observations.

\subsection{Data Acquisition}
Signal sampling is triggered by an opto-interrupter which is switched by a 64-tooth wheel physically attached to the wave plate.   As a result, the sampled data is stored in a coordinate system fixed to the wave plate position, rather than as a timestream, simplifying data analysis.  For a physical waveplate rotation rate $f_{wp} \sim 0.15$ Hz, this allows us to sample the bolometers at $\sim 20$ Hz, well above the signal frequency of $4 f_{wp} = 0.6$ Hz and a factor of several above a detector bandwidth-defining low-pass filter at 6 Hz.  The signal frequency 0.6 Hz also lies well above the 10 mHz $1/f$ knee of the readout electronics.  The dark bolometers and two thermistors used to monitor the temperature of the bolometer stage are also sampled at $\sim 20$ Hz.  Basic cryogenic and other housekeeping data are sampled at a lower date rate.  The data is stored locally on a ruggedized portable computer located on the telescope, then transferred via FTP to our data analysis computer nightly. 

The experiment is designed to run autonomously for weeks at a time.  It is controlled by UNIX-based scheduling software installed on a PC in a shed near the telescope.  Commands are sent by the scheduler to either the telescope control computer (a VAX system) or to the portable computer.  The portable computer runs Windows and LabVIEW.  Upon receipt of commands from the scheduler, the LabVIEW software carries out such tasks as cycling the $^{3}$He refrigerator for the bolometer stage, taking and storing data, and starting and stopping the wave plate rotation.  The LabVIEW software also returns some data to the scheduler, so that the scheduler can automatically carry out tasks such as pointing, skydips, and calibration.

\subsection{Anticipated Sensitivity}
Four main sources of noise limit the ultimate sensitivity of the Polatron:  background photon noise; detector noise; amplifier noise; and fluctuations in atmospheric emission.  We will estimate the contributions from each of these below; the quantitative results are summarized in Table 2.  It is conventional to express the noise in terms of Noise Equivalent Power (NEP), measured in units of W Hz$^{-1/2}$.

Photon noise arises from quantum fluctuations in emission from the atmosphere and receiver environment.  The photon noise equivalent power is proportional to the square root of the number of photons observed, which is in turn proportional to the optical power incident on the detectors.  The optical power is dominated by in- and out-of-band emission from the 118 GHz O$_{2}$ line, which remains fairly constant over changing weather conditions.  The most variable component is 60 GHz H$_{2}$O line emission, which is quantified by the precipitable water vapor column density (\textit{ppwv}).  Between September and May, \textit{ppwv} for the OVRO site typically ranges between 3 and 8 mm (see \url{www.ovro.caltech.edu} for a snapshot of the current and 30-day weather conditions).

In the direction of the North Celestial Pole, we estimate a background loading temperature of 60 K for our spectral bandpass.  The optical loading on each bolometer is 
\begin{equation}
P_{opt} = \int_{0}^{\infty} d\nu \frac{\eta(\nu)}{2} (A \Omega) B_{\nu}(T_{load}), 
\end{equation}
where the factor of $1/2$ accounts for the split in polarization, $\eta(\nu)$ is the spectral efficiency, $A \Omega = \lambda^{2}$ is the fixed throughput of the single mode feed horn, and $B_{\nu}$ is a thermal spectrum.  In the limit of low effective photon occupation number ($\eta_{opt} < e^{h\nu/kT_{load}}-1$), the photon NEP is  
\begin{equation}
NEP_{photon} \approx \sqrt{P_{opt}h \nu_{0}},
\end{equation}
where $h \nu_{0}$ is the average photon energy \citep{richards94}.  This is per detector, so the total noise will be increased by $\sqrt{2}$ upon differencing the two uncorrelated detectors.
  
The second component of noise is fundamental thermal fluctuations in the detectors.  A bolometer consists of an absorber, in this case an open mesh of silicon-nitride in a spider-web shape, coated with a layer of gold; and a thermistor, which is typically a semiconductor chip with temperature-dependent resistance.  The absorber and thermistor are suspended at temperature $T$ from a thermal bath $T_{0}$ by some thermal conductivity $G$.  Detector noise arises as phonon shot noise in the thermal link between the thermistor and the bath or due to Johnson noise in the thermistor \citep{richards94}.  For our operating configuration, the phonon noise dominates:
\begin{equation}
NEP_{detector} \approx \sqrt{4 k_{B} T^{2} G},
\end{equation}
where $k_{B}$ is the Boltzman constant.  As above, this noise will increase by $\sqrt{2}$ upon differencing.

The third component of noise is amplifier noise in the cold matched JFETs.  The measured differential noise performance of $\sim 4$ nV Hz$^{-1/2}$ is converted to estimated NEP by dividing by an estimate of the bolometer responsivity under the observing conditions expected at the telescope.  

The final component of noise is $1/\textit{f}$ fluctuations of atmospheric water vapor content.  Differencing the two detector signals rejects common-mode fluctuations.  This technique is identical to that used in the Sunyaev-Zel'dovich Infrared Experiment (SuZIE), except that SuZIE differences the power in two spatial pixels, whereas the Polatron differences the power in two senses of linear polarization within the same spatial pixel.  Since the atmosphere is not significantly polarized \citep{keating98}, atmospheric $1/\textit{f}$ noise is limited only by the common mode rejection ratio (CMRR) of the receiver.  Ground-based measurements of CMB polarization would not be feasible without cancellation of atmospheric fluctuations in this manner.

The amount of residual atmospheric emission fluctuations can be estimated using the methods of \citet{church95}.  Since SuZIE and the Polatron have similar beam sizes, the atmospheric fluctuations observed by SuZIE at 217 GHz from Mauna Kea are scaled to obtain an estimate of atmospheric fluctuations that will be seen by the Polatron at 100 GHz from OVRO.  This scaling relies on the following assumptions:  that residual atmospheric noise is dominated by gain mismatches and not beam mismatches; that Zeeman splitting of O$_2$ lines by the Earth's magnetic field is negligible; that the gains of the two polarization channels can be matched on a regular basis to 1\%; that fluctuations are due to variations in water vapor content; and that the OVRO atmospheric fluctuations can be estimated by frequency scaling Mauna Kea fluctuations as measured by SuZIE.  We calculate
\begin{eqnarray}
NEP_{atmos} & = & \frac{NEP_{S}}{CMRR} \times R_{1} R_{2} \times \left(\frac{A\Omega_{P}}{A\Omega_{S}}\right) \left(\frac{\nu_{P}}{\nu_{S}}\right)^{2} \nonumber \\
            & & \times  \left(\frac{(\Delta\nu / \nu)_{P}} {(\Delta\nu / \nu)_{S}} \right) \left(\frac{\eta_{P}/2}{\eta_{S}}\right).
\end{eqnarray}
Here, the subscript $P$ refers to the Polatron and $S$ refers to SuZIE.  The quantity NEP$_{S}$ is the atmospheric NEP = 1.0 $\times 10^{-15}$ W Hz$^{-1/2}$ measured by SuZIE at Mauna Kea in a single channel at the Polatron observing frequency $4f_{wp} = 0.6$ Hz.  R$_{1}$ encodes the difference in atmospheric opacities at the two sites:
\begin{equation}
R_{1} =  e^{-\frac {\tau_{S} - \tau_{P}} {\sin(\theta)} },
\end{equation}
where the zenith opacity $\tau_{S} = 0.07$ at Mauna Kea during the SuZIE measurement, $\tau_{P} \sim 0.5$ at OVRO, and the assumed observation angle is $(\frac{\pi}{2} - \theta) \sim 30^{\circ}$.  R$_{2}$ = 0.18 is a scaling factor which takes into account the difference in the contribution to the total atmospheric opacity due to water vapor at $\nu_{P}$ = 98 GHz and $\nu_{S}$ = 217 Ghz (see \citet{danese89}).  A factor of 1/2 is included to convert the SuZIE data to a single polarization.

The total noise performance of the system is predicted by adding the NEP from each source in quadrature:
\begin{eqnarray}
NEP_{total}^{2} & = & NEP_{photon}^{2} + NEP_{detector}^{2} + NEP_{amplifier}^{2} \nonumber \\
                & & + NEP_{atmos}^{2}.
\end{eqnarray}

The NEP is converted to Noise Equivalent CMB Temperature NET$_{cmb}$, measured in $\mu$K$s^{1/2}$, as follows.  First, we calculate the Noise Equivalent Flux Density (NEFD), measured in Jy~s$^{1/2}$, which is the flux sensitivity of the receiver in a single second of integration time.  As in (17) above,
\begin{equation}
NEP_{total} \approx \Delta\nu \frac{\eta}{2}\,  A \times NEFD \times \sqrt{2}.
\end{equation}
The factor of 1/2 indicates that we are calculating the sensitivity to a polarized signal, $A$ is the illuminated area of the telescope primary, $\eta \equiv \overline{\eta(\nu)}$ is the band-averaged optical efficiency of the receiver, $\Delta\nu$ is the spectral bandwidth, and the final factor of $\sqrt{2}$ accounts for the conversion from Hz$^{-1/2}$ to s$^{1/2}$.  The actual CMB flux per beam, $F_{cmb}$, measured in Jy, is 
\begin{equation}
F_{cmb} \approx \frac{2h\nu_{0}^3}{c^2}\frac{1}{e^{x}-1}\times \Omega,
\end{equation}
where $\Omega$ is the beam solid angle and $x = h\nu_{0}/kT_{cmb}$.  Converting from intensity to CMB temperature, the NET$_{cmb}$ is then
\begin{equation}
\frac{NET_{cmb}}{T_{cmb}} = \frac{e^{x}-1}{xe^{x}}\times\frac{NEFD}{F_{cmb}}.
\end{equation}
Calculated NEFD and NET$_{cmb}$ for different atmospheric conditions are included in Table 2.

\subsection{Calibration}
Calibration of the absolute responsivity of the instrument is accomplished through single-polarization observation of largely unpolarized sources, such as planets.  Each single-channel signal will be dominated by the atmospheric fluctuations described above.  As before, the level of such fluctuations observed with the SuZIE receiver can be scaled to predict noise levels for the Polatron, albeit with no common-mode rejection and no advantage due to the polarization modulation at 4$f_{wp}$.  A typical drift scan over a calibrator occurs over a time period of $\sim 30$ seconds.  We expect a single-channel NEFD on those time scales of $\sim$ 140 Jy s$^{1/2}$.  The anticipated flux from a typical calibrator (Mars, for instance) is on the order of several hundred Jy, so several drifts over the same source may be required in poor weather.  Hence, calibration uncertainty will be dominated, as it is in the SuZIE experiment, by the uncertainty in the brightness of the calibrators themselves.  We expect to flux calibrate to the 5-10\% accuracy typically achieved for CMB observations. 

\subsection{Systematic Polarization}
Until recently, millimeter-wave polarimetry has consisted mainly of the study of magnetic fields through the measurement of the polarization patterns of molecular clouds and compact dusty sources.  The quantity of scientific interest for these studies has been the \textit{degree of polarization} (or, simply, \textit{polarization}) of a source, a vector-like quantity which has an alignment direction $\theta_{source}$, modulo rotation by $180^{\circ}$, and positive fractional magnitude $0 < p_{source} < 1$  for every observed position $(\theta, \phi)$ on the sky.  This quantity is often denoted $\overline{p}_{source} (\theta,\phi)$; maps of this pseudo-vector field are often overlaid on emission intensity contours so that the relationship between the magnetic field and the emissive structure of the source is demonstrated.

Typical dust sources are polarized at the $1 - 10\%$ level, so a clean measurement of such should be accurate to $< 1\%$.  A useful quantity to define is the fractional \textit{systematic polarization}, $\overline{p}_{sys}$, of a system, which is the percentage polarization measured when an intrinsically unpolarized source is observed.  By definition this quantity does not vary with source brightness.  It is usually caused by, and fixed with respect to, the telescope and receiver.  It can be measured through observation of a known unpolarized source over a period of time such that the parallactic angle of the source rotates with respect to the telescope.  The polarization as a function of time measured in the telescope's coordinate frame should describe a circle with radius $p_{source}$ and center offset by $\overline{p}_{sys}$ from the origin.  Once $\overline{p}_{sys}$ is measured in this way it can be simply subtracted pixel by pixel from maps of observed polarization $\overline{p}_{obs} (\theta,\phi)$:
\begin{equation}
\overline{p}_{source} (\theta,\phi) = \overline{p}_{obs} (\theta,\phi) - \overline{p}_{sys}.
\end{equation}

In the low signal-to-noise regime, polarization percentage can be a difficult quantity to analyze, since it is not Gaussian-distributed about a mean of zero.  To remedy this, reduced (fractional) Stokes parameters $q = Q / I$ and $u = U / I$ are introduced, with $Q$ and $U$ the measured Stokes parameters and $I$ the measured source intensity.  The quantities $q$ and $u$ can be positive or negative.  The systematic polarization adds in the same way as before:
\begin{eqnarray}
q_{source}(\theta,\phi) & = & q_{obs}(\theta,\phi) - q_{sys}, \\
u_{source}(\theta,\phi) & = & u_{obs}(\theta,\phi) - u_{sys}.
\end{eqnarray}
The polarization pattern can still be recovered since $p_{source} = \sqrt{q_{source}^2 + u_{source}^2}$ and $\theta_{source} = \frac{1}{2}\tan^{-1}(U / Q)$.

For measurements of bright, polarized sources such as dust and molecular clouds, this characterization of systematic errors relating to polarization is adequate.  CMB polarization measurements, however, are different.  The non-normalized Stokes parameters $Q$ and $U$ are of interest, rather than the fractional polarization $p$ or the reduced parameters $q$ and $u$.  The underlying intensity distribution is bright, with temperature 2.728 K, and extremely uniform, but varying across the sky with some likely unknown amplitude $\Delta T / T \sim 10^{-5}$.  Hence, given a measurement of systematic polarization $q_{sys}$ from observation of a polarized source, the resultant systematic $Q_{sys}(\theta,\phi) = q_{sys}T(\theta,\phi)$ cannot be simply calculated and subtracted from the data, since $T(\theta,\phi)$ is, as yet, unknown.  Since we are looking for fluctuations $\Delta Q / T \sim 10^{-6}$, we require $q_{sys}$ and $u_{sys} \lesssim 0.01$ to ensure that we are not systematically polarizing $\Delta T$ fluctuations.  An observing strategy which measures the polarization of a single pixel at different parallactic angles allows this systematic polarization to be subtracted to $< 1\%$ even if the underlying intensity distribution is unknown.
 
Signal pick-up in the far sidelobes of the main beam can affect faint-background measurements if the source in the sidelobes is bright enough.  The sun, the moon, the galaxy, and the horizon are examples of such sources.  In addition, for polarization measurements, the two components of linear polarization will have different beam shapes, since the polarization analyzer (in our case the OMT) defines a set of coordinates with respect to the receiver, telescope, and telescope surroundings.  The differential pick-up from the source in the two beams should not be greater than the expected CMB polarization signal.  The observation strategy, then, is subject to a certain set of restrictions on the angle between the observed region of sky and these various sources.  Observation near the North Celestial Pole mitigates all of these effects.

\subsubsection{Systematic Polarized Flux}
In contrast to the systematic polarization described above, which amounts to a \textit{percentage} polarization acquired by an observed unpolarized source, the Polatron will observe a variety of systematic polarized fluxes which are independent of the brightness of an observed source.  For instance, thermal radiation from the surrounding environment will scatter from the telescope feed legs into our system, picking up some polarization during the scattering.  This type and most types of polarized flux will vary slowly, and will be removed from the data as a DC-level while drift-scanning.  While actively scanning the telescope, spurious signals can be removed by referencing them to the telescope's surrounding environment, while astrophysical sources will rotate about the North Celestial Pole.

The wave plate itself is a source of systematic polarized flux.  As a consequence of the method used to produce birefringence, the two crystal axes of the wave plate incur differing loss tangents.  Unpolarized sky radiation incident on the plate will be converted to a systematic flux.  This signal, however, will be modulated at $2f_{wp}$ and will not contribute to the signal after lock-in at $4f_{wp}$.  

\subsubsection{Receiver Polarization}
Gain mismatches between the two bolometer channels generate a fractional systematic polarization such that an unpolarized source of intensity $I(t)$ incident on the receiver will produce a signal $Q'_{rec}(t) = q'_{rec}I(t)$ which is \textit{not} modulated by the wave plate at frequency $4f_{wp}$.  The magnitude of the receiver polarization is inversely related to the common mode rejection ratio of the instrument:  $|q'_{rec}(t)| = 1 / CMRR$.  If the input intensity $I(t)$ varies on time scales $t \sim 1 / 4f_{wp}$, then the demodulated signal at $4f_{wp}$ will include components of $I(t)$.  An example of this effect was treated earlier ($\S2.5$) for the case of unpolarized sky fluctuations.  As was shown, the receiver polarization can be reduced to beneath $1\%$ by adjustment of the relative gains of the two channels.  

Another source of common-mode signal is variation in the temperature of the bolometer stage.  Variations at frequency $4f_{wp}$ will be recorded as signal.  Because the gain differs for optical signals incident on the receiver as opposed to thermal signals at the bolometers themselves, a specific relative gain adjustment which trims out common-mode optical signals is not likely to also trim out common-mode temperature fluctuations.  For this reason the bolometer stage will be temperature controlled to $\sim 100$ nK Hz$^{-1/2}$ stability using methods described in \citet{holzapfel97}.

\subsection{Astrophysical Foregrounds}
The Polatron spectral band is close to the minimum of contamination estimated to arise from observing through our own galaxy.  Polarized foreground processes include synchrotron emission, free-free emission, thermal dust emission, and, perhaps, emission from spinning dust grains. 

Galactic synchrotron emission is likely to be highly polarized.  \citet{bouchet98} point out that spectral index and polarization information obtained from long-wavelength observations of galactic synchrotron might not extend to millimeter wavelengths.  Nevertheless, by making the assumptions that the emission is 44\% polarized and has the same spatial distribution as the unpolarized emission at long wavelengths, they predict at 100 GHz the following 
angular power spectra between galactic latitudes $30^{\circ}$ and $75^{\circ}$:
\begin{eqnarray}
C_{G}^{synch} & = & 0.9 ~l^{-3} ~{\mu K}^2 \\
C_{C}^{synch} & = & 0.9 ~l^{-3} ~{\mu K}^2
\end{eqnarray}

Observations of galactic dust regions at 100 $\mu m$ indicate that where dust emission is bright, the level of polarization is of order 2\% \citep{hildebrand95} apart from a few small regions in which the polarization rises to 10\%.  However, the degree of polarization of high latitude cirrus emission is unknown.  A model for high-latitude galactic dust polarized emission has been created by \citet{prunet00}.  This model assumes that cirrus is traced by observed \ion{H}{1} emission and that the dust grains are similar in shape to those observed by \citet{hildebrand95}.  A 17.5 K Planck spectrum with emissivity proportional to $\nu^{2}$ is assumed.  The dust grains are expected to align with the galactic magnetic field; an unfavorable orientation for the magnetic field is used.  The assumed intrinsic dust polarized emissivity of $\sim 30\%$ is reduced due to projection effects.  At 100 GHz, the angular power spectra predicted by \citet{bouchet98} between galactic latitudes $30^{\circ}$ and $75^{\circ}$ are
\begin{eqnarray}
C_{G}^{dust} & = &~8.9 \times 10^{-4} ~l^{-1.3} ~{\mu K}^2 \\
C_{C}^{dust} & = & 10.0 \times 10^{-4} ~l^{-1.4} ~{\mu K}^2
\end{eqnarray}

Free-free emission in \ion{H}{2} regions, like the CMB, can pick up some small polarization through Thomson scattering.  This is likely to generate a level of polarization much lower than that of synchrotron radiation.  Likewise, the spinning dust grains proposed by \citet{draine98} should not be polarized at or near the level of synchrotron emission.  Neither of these is likely to contaminate our measurements.  The galactic synchrotron and dust polarized emission at 100 GHz is more than an order of magnitude below the CMB signal over the $l$-range to which the Polatron is sensitive (see Figure 3).

\subsection{Observing Strategy}
The choice of observing strategy for the Polatron is dictated by numerous practical observing constraints, such as limited observing time and telescope design. For a CMB polarimeter, the observing strategy should provide the capability to detect both Stokes parameters $Q$ and $U$. All three polarization power spectra, $C^{GG}_\ell,C^{CC}_\ell$ and  $C^{GT}_\ell$, depend on the polarization tensor $P_{ab}$ (see (\ref{pab})), which is a function of both $Q$ and $U$.  Without measurement of both $Q$ and $U$ it is therefore impossible to obtain model-independent detections of the power  spectra needed to statistically characterize the CMB, and ultimately, to draw conclusions on the underlying cosmological parameters of interest. The Polatron will simultaneously measure both $Q$ and $U$, and is therefore able to detect all three polarization power spectra in a model-independent fashion. In this section, we reduce the available  phase-space of experimental observing strategies to a tractable number of parameters which can subsequently be systematically investigated.

\subsubsection{Observing Parameters}

The Polatron will measure the polarization of the CMB in two operational modes. Initially, we will observe in the `initial detection' mode, in which the goal of the experiment is to unambiguously detect the polarization signal.  Following detection we will enter the `$\ell$-space spectroscopy' mode, in which the goal is to measure the angular power spectra $C^G_\ell$ and $C^C_\ell$. In this section we primarily concentrate on the former observing mode, but will address concerns relevant to the latter as well.

The 5.5 m telescope mount at OVRO is a conventional altitude-azimuth platform most recently used in a study of the Sunyaev-Zeldovich effect in galaxy clusters \citep{mason01}. The size of the telescope prevents the implementation of ground screens typically used to minimize the effect of ground spillover. Our first level of sidelobe contamination mitigation is provided by an  edge taper of -20 dB. In addition, we will restrict the telescope's elevation to a single zenith angle and minimize the azimuthal scan amplitude of the telescope. These constraints motivate us to scan rings about the North Celestial Pole (NCP), which also gives us the ability to observe in an anti-solar direction during daylight observations. In the following subsection we consider the optimization of the ring opening angle.

\subsubsection{Using CMB Anisotropy as a Prior}
At present, there exist no detections of the polarization of the CMB. However, our pursuit of an initial detection can be guided by the numerous robust detections of CMB anisotropy that appear in the literature (see, for example, \citet{lange00}). The strong correlation between CMB temperature anisotropy and polarization \citep{zaldarriaga97, hu97} provides one method for optimizing a CMB polarimeter's observation strategy. In addition, given a high signal-to-noise temperature anisotropy map, temperature-polarization cross-correlation may be easier to detect than the polarization auto-correlation \citep{crittenden95}.

Given a well sampled temperature map one can determine regions where the temperature-polarization cross-correlation is expected to be large. These regions are \emph{not} at the location of hotspots and coldspots in the anisotropy as the cross-correlation function vanishes at zero-lag due to isotropy considerations.  Rather, maximum correlation occurs in regions where the photon-baryon fluid's velocity at last-scattering is large, which roughly translates into regions where the CMB anisotropy displays a large spatial gradient.

In addition to the cross-correlation approach, which is dependent on an underlying CMB temperature anisotropy map to correlate with, the second use of temperature anisotropy detections is statistical.  As the generation of CMB polarization is a causal process which is completely dependent on the CMB anisotropy, there is an intimate relationship between the spectral behavior of polarization and anisotropy. By designing our observing strategy carefully, using the anisotropy power spectrum as a guide, we can maximize our chances for a positive detection of CMB polarization.

\subsubsection{CMB Polarization: Initial Detection Observing Mode}

For the initial detection observing mode, we place control over systematic effects at a premium. Therefore, we are led to consider observing strategies that produce the most robust detection of the variance or amplitude of the polarization power spectrum. The cost of this approach is that we lose details of the precise shape of the spectrum. This approach is justified for an experiment, such as the Polatron, which seeks to make an initial detection of CMB polarization.

In order to obtain the highest signal-to-noise ratio per pixel in a fixed amount of observing time, we should restrict our sky coverage to a small number of pixels. It is also advantageous to fix the elevation angle of the telescope to eliminate differential ground spillover from multiple elevation pointings.  These factors motivate us to observe rings about the NCP of radius $\theta_o$. This could be accomplished in several different ways.  The simplest method is to fix the telescope in elevation and execute a drift scan in this position for the duration of the observing campaign. Each pixel on the ring is observed only once per day and is only referenced to pixels that are adjacent to it along the ring. This mode of observation is thus susceptible to the effects of long time-scale correlations between pixels caused by low frequency drifts in the instrument response, which would lead to time-variable offsets in the final map reconstruction \citep{tegmark97}.

A preferable observing mode is to fix the telescope in elevation, $\theta_o$, and ``nod'' in azimuth by an amount $\Delta \phi$ across the North Celestial Pole at a frequency higher than the residual $1/f$ knee of the instrumental response after lock-in at the wave plate rotation frequency. The Stokes parameters at $(\theta_o,\phi)$ are differenced from those at $(\theta_o,\phi + \Delta \phi)$ which removes residual correlated noise occurring at timescales longer than the ``nod'' timescale. A similar strategy was chosen for the Ring 5m and Ring 40m campaigns \citep{leitch00, myers93}, and has the virtue that the telescope is never moved in elevation, which reduces the modulated side lobe power entering the receiver.

To analyze this observing mode we use the small-angle, flat-sky limit, which is justified for $\theta_o \leq 10\arcdeg$. We will follow \citet{zaldarriaga98} and \citet{delabrouille98} and work with Fourier transforms instead of spherical
harmonic transforms. The data set is a map of $Q_j$ and $U_j$, measured at $N$ pixels along the ring, indexed by position angle along the ring $\phi_j = 2\pi\,\frac{j-1}{N}$. The maps of $Q_j$ and $U_j$ are spectrally decomposed into Fourier modes,
\begin{eqnarray}
Q^k & = & \frac{1}{N}\sum_j Q_j e^{-ik\phi_j} \nonumber \\
    & = & -\sum_{\ell \geq \vert k \vert} \sqrt{(2\ell+1)/4\pi} B_{\ell k}[a^{G}_{\ell k} F_{1,\ell k} (\theta_o)+ i a^{C}_{\ell k} F_{2,\ell k} (\theta_o)] \nonumber \\
    & & + N^k_Q, 
\end{eqnarray}
and
\begin{eqnarray} 
U^k & = & \frac{1}{N}\sum_j U_j e^{-ik\phi_j} \nonumber \\
    & = & -\sum_{\ell \geq \vert k \vert} \sqrt{(2\ell+1)/4\pi} B_{\ell k}[a^{C}_{\ell k} F_{1,\ell k} (\theta_o) - i a^{G}_{\ell k} F_{2,\ell k}(\theta_o)] \nonumber \\
    & & + N^k_U, 
\end{eqnarray}
where $B_{\ell k}$ is the Fourier transform of the instrumental beam,  $N^k_Q, N^k_U$ are the contributions to the experimental noise in the $k$-th mode of $Q,U$, such that $\lan N^k_{S}N^{k'}_{S'} \ran = w^{-1}\delta_{k,k'}\delta_{S,S'}$, where $w^{-1}=\sigma^2/N =  NET^2/t_{obs}$,  with $\sigma$ being the pixel noise, $NET$ being the noise equivalent temperature (defined in section 2.5) measuring each Stokes parameter $S,S'  \in \{ Q,U \}$, and $t_{obs}$ is the total observation time. The functions $F_{1,\ell k},F_{2,\ell k}$ account for the behavior of the Stokes parameters under rotations on the celestial sphere and can be written in terms of ordinary Legendre polynomials \citep{kamionkowski97}. These functions, in combination with the $B_{\ell k}$, determine the ``window'' in multipole space to which each Fourier mode, $k$, is sensitive.

The correlation matrix of the mode expansion of the Stokes parameters is \citep{zaldarriaga98},
\begin{eqnarray}
\lan Q^{k}Q^{k'}\ran & =& \delta_{k,k'}\sum_{\ell} (2\ell + 1)/4\pi (W_{1,\ell k} C_{G,\ell} + W_{2,\ell k} C_{C,\ell}) \nonumber \\
& & + \delta_{k,k'} w^{-1},
\end{eqnarray}
and similarly for
\begin{eqnarray}
\lan U^{k}U^{k'}\ran & =& \delta_{k,k'}\sum_{\ell} (2\ell + 1)/4\pi (W_{2,\ell k} C_{G,\ell} + W_{1,\ell k} C_{C,\ell}) \nonumber \\
&& + \delta_{k,k'} w^{-1}, 
\end{eqnarray}
where the window functions $W_{1,\ell k} = B^2_{\ell k} F^2_{1,\ell k}$ and $W_{2,\ell k} = B^2_{\ell k} F^2_{2,\ell k}$ determine the contribution of the $\ell$-th mode of the power spectra to a given ring $k$-mode.

Following \citet{zaldarriaga98}, we now obtain approximations for the window functions in the small-scale (flat-sky) limit. The maps can be expanded in two dimensional Fourier transforms as,
\begin{eqnarray}
Q(\theta_o,\phi) & = & (2\pi)^{-2}\int d^2\ell
e^{i\ell\theta_o\cos(\phi-\phi_\ell)} [G(\vec{\ell})
\cos{2(\phi-\phi_{\ell})}  \nonumber \\
& & - C(\vec{\ell})\sin{2(\phi-\phi_{\ell})}],
\end{eqnarray}
and
\begin{eqnarray}
U(\theta_o,\phi) & = & (2\pi)^{-2}\int d^2\ell
e^{i\ell\theta_o\cos(\phi-\phi_\ell)} [G(\vec{\ell})
\sin{2(\phi-\phi_{\ell})}  \nonumber \\
& & + C(\vec{\ell})\cos{2(\phi-\phi_{\ell})}].
\end{eqnarray}
Here, $G(\vec{\ell})$ and $C(\vec{\ell})$ are the two dimensional Legendre transforms of the grad and curl fields, respectively, and $\vec{\ell}$ indexes the mode with polar representation $\vert \ell \vert e^{i\phi_\ell}$. The statistical properties of the grad and curl fields are defined by 
$\lan
G(\vec{\ell_1})G(\vec{\ell_2})\ran = (2\pi)^{2}C^G_\ell \, \delta
(\vec{\ell_1}-\vec{\ell_2})$, $\lan
C(\vec{\ell_1})C(\vec{\ell_2})\ran = (2\pi)^{2}C^C_\ell \, \delta
(\vec{\ell_1}-\vec{\ell_2})$, and $\lan
C(\vec{\ell_1})C(\vec{\ell_2})\ran = 0$. 
Fourier transforming from $(\theta_o,\phi)$ space to $k$ space and calculating the correlation matrix for the $k$ modes yields,
\begin{eqnarray}
\lan Q^{k}Q^{k'}\ran & =& \delta_{k,k'}\int\frac{\ell d\ell}{2\pi}\, \, \{ C^{G}_{\ell}\onequarter  \, [J_{k+2} (u) +
J_{k-2}(u)]^2 \nonumber \\
& & + C^{C}_{\ell} [J_{k+2} (u) - J_{k-2}(u)]^2\},
\end{eqnarray}
and similarly,
\begin{eqnarray}
\lan U^{k}U^{k'}\ran & =& \delta_{k,k'}\int\frac{\ell d\ell}{2\pi}\,\onequarter \, \{C^{C}_{\ell}  \, [J_{k+2} (u) -
J_{k-2}(u)]^2  \nonumber \\
& & + C^{G}_{\ell}[J_{k+2} (u) + J_{k-2}(u)]^2\}.
\end{eqnarray}
The window functions in the small-scale limit are therefore given by:
\beq \label{eq:windows1} W_{1,lk} = \onequarter B_{\ell k}^2[J_{k+2} (u) +J_{k-2}(u)]^2, \eeq
and \beq \label{eq:windows2}W_{2,lk} = \onequarter B_{\ell k}^2[J_{k+2} (u) - J_{k-2}(u)]^2 .\eeq

The noise per pixel is assumed to be independent of pixel and uncorrelated between $Q$ and $U$, which, in terms of the noise per mode (also uncorrelated between $Q$ and $U$) leads to $$ \langle {\bar N}_Q^{k*}{\bar N}_Q^{k^\prime} \rangle = w^{-1} \delta_{k,k^{\prime}}, $$ where $w^{-1} \equiv \sigma^2 /N_{pix}$, with $\sigma$ being the pixel noise.

Given a theoretical model (characterized by its temperature and polarization angular power spectra) we calculate the sum of the signal-to-noise eigenmodes for both $Q$ and $U$. This is repeated for a range of observing strategies characterized by the opening angle of the ring. We utilize CMBFAST (\url{www.physics.nyu.edu/matiasz/CMBFAST/cmbfast.html}) to generate polarization power spectra from a cosmological model that is compatible with the results of several recent temperature anisotropy measurements \citep{lange00}. The grad-mode power spectrum is shown in Figure 1, and the cosmological parameters are indicated in the caption of that figure. Though not a true constrained realization (which would utilize the polarization temperature cross-correlations as well), this method provides a useful first-order approximation to the
true underlying polarization power spectrum. \citet{jaffe00} have made a similar calculation for observations of square shaped regions on the sky and find that the optimal sky coverage is rather model-independent for a fixed $\Omega_b$. We
expect that our choice of fiducial model will allow us to optimize the sky coverage for the Polatron as well. \citet{jaffe00} also calculate the optimal sky coverage as a function of beam size. The detectability (signal to
noise ratio) of the gradient mode signal is found to be strongly dependent on the beam-size, $\theta_{fwhm}$, for $0.1\arcdeg < \theta_{fwhm} < 1.0 \arcdeg$. In this regime, smaller beams are found to produce the highest signal to noise detections. For $\theta_{fwhm}<0.1\arcdeg$ the dependence on beam-size is much weaker, as nearly all of the power in the gradient mode signal is at scales greater than $0.1\arcdeg$, corresponding to multipoles $\ell< 2000$.

Figure 5 displays the results of the optimization for the ring observing strategy, assuming three weeks of continuous
observations. The dotted line displays the performance of a hypothetical experiment with no $1/f$ noise; we see that the
expected signal-to-noise ratio is optimized when $\theta \simeq 0.1 \arcdeg$. This can be translated into a constraint on the number of beam sized pixels in the map, $N_{pix} \sim 10$.

This result can be understood in the initial detection mode as a desire to reduce the noise variance until a detection is made, which translates into deep integration on a small number of pixels. Once a detection is made, the area of sky observed should be increased to reduce the effects of sample (or ``cosmic'') variance. This will be treated in the following subsection.

We note here that two effects can alter the calculation of the optimal ring angle: foregrounds and $1/f$ noise. Little is known about the nature of polarized foregrounds. Heuristically, if polarized foregrounds follow a falling spectrum $\sim 1/\ell^{\beta}$ (\emph{e.g.} \citet{prunet00}, and $\S 2.8$), they will be correlated over large angles on the sky, with a characteristic size $\Theta_{fore}$. In order to subtract the foreground with high precision, we must accurately characterize its correlation properties (or its power spectrum).  To reduce the sample variance of the foreground measurement, we are driven towards observing a region that contains many pixel pairs separated by a variety of scales. For a ring observing geometry, increasing the ring diameter $\theta_{0} > \Theta_{fore}$ increases the number of pixel pairs, which leads to a higher precision characterization of the foreground.

A similar effect that also drives us to increase the ring opening angle over the formally optimal value is the effect of $1/f$ or ``pink" noise. In the presence of $1/f$ noise, the overall signal-to-noise ratio decreases and the optimum ring angle increases significantly as shown in the dashed curves of Figure 5 to $> 1 \arcdeg$. Residual $1/f$ noise increases the noise of the lowest spatial frequency modes in $k$ space. The signal at a scale $k$ is reduced by an effective $k$-space window that falls as $w_k = k^{-1}$. This makes it advantageous to remove the lowest $k$ modes from the survey. This can be accomplished quite naturally by ``nodding'' the telescope in azimuth between two points on the ring separated by $90\arcdeg$ or 6h in Right Ascension. This removes the $k=2$, or quadrupole, component of the signal power spectrum. Translated back into $\ell$ space, this strategy effectively acts as a high-pass filter on $C^G_\ell$ and $C^C_\ell$.  The signal-to-noise curve for this differencing scheme is indicated by the solid line of Figure 5, where it is shown that $\theta \simeq 1 \arcdeg$ yields the optimum signal-to-noise ratio. The nodding allows the low frequency $k$-modes of the power spectra to survive at the cost of the quadrupole component. Although the quadrupole is usually the dominant $k$-space multipole for a ring experiment \citep{zaldarriaga98}, for a realistic experiment, with $1/f$ noise, a comparison of the solid and dashed curves of Figure 5 demonstrates that the  quadrupole-differencing strategy is clearly preferred. The nodding of the telescope, combined with the modulation of the Stokes parameters via the rotation of the waveplate, makes our observing strategy similar to a ``double-switched'' scheme, commonly used in temperature anisotropy measurements.

The effective window function for the combination of all modes (except $k=2$) is shown in Figure 1 along with the polarization power spectra for our fiducial model. The window function for the combined modes displays the low-pass behavior noted above, and is rather broad in $\ell$-space, which is caused by the narrowness of the ring along the radial direction. Since the map is only one pixel wide in radius, only one radial mode is present in the 2-dimensional Fourier transform of the data.  This violates the conventional rule of thumb that suggests that the map dimensions
should be comparable in radius and azimuth \citep{tegmark97}. A ring observing strategy may also complicate the separation of gradient and curl modes in the final power spectrum estimates \citep{tegmark00}. However, the ring observing strategy is capable of providing an initial, unambiguous, statistical detection of the variance of the polarization. Since the ring strategy only provides information regarding the amplitude of the angular power spectrum, after detection we will switch to an observing strategy that can detect its shape. This could be realized by combining several rings at different declinations to form a two-dimensional cap or annulus. The simplicity of implementation, redundancy of information, and minimization of systematic effects make the ring strategy a conservative approach for an instrument that is seeking to make an initial detection of CMB polarization.

The optimum ring opening angle also depends on the integration time of the observing campaign for a fixed detector noise.  Integrating on a small region of the sky for the duration of the campaign will reduce the pixel-noise variance, but will increase the sample variance of the observation since fewer pixels are observed. The optimum number of pixels to observe thus depends on the signal-to-noise regime of the experiment. For short observations, it is better to observe fewer pixels until a detection is made and then increase the number of pixels measured to this same sensitivity. Figure 5 assumes a constant detector noise, with the total observation time increasing between the three panels. The left panel shows the optimal number of pixels for a short integration (3 weeks) where we find only $\sim 10$ pixels should be observed. The middle panel shows an intermediate-length campaign (18 weeks) where a $\sim 2 \sigma$ detection would occur for the quadrature-differenced data, even in the presence of $1/f$ noise. The right-most panel shows a long
integration (36 weeks) resulting in a robust detection with a large number of pixels. This indicates that, as the noise
variance decreases, we should enlarge the area of sky observed in order to reduce the sample-variance contribution. Ultimately, this leads us to cover a 2-dimensional patch of sky following an initial detection.

For the initial-detection mode we choose a ring opening angle of $\theta = 1.8\arcdeg$ motivated by these considerations and the desire to observe fields which have also been observed by the previous successful CMB anisotropy observing campaigns conducted at OVRO: OVRO NCP \citep{readhead89}, Ring 40m \citep{myers93} and Ring 5m \citep{leitch00}. The Ring 40m 20 GHz data were taken on a ring at declination $\delta = 88\arcdeg 10\arcmin 42\arcsec$ which corresponds to an opening angle of $\theta \simeq 1.82 \arcdeg$ with a resolution of $1.8\arcmin$. The Ring 5m data, taken with the same telescope as the Polatron at 30 GHz, with a resolution of $\simeq 7\arcmin$ on the $\delta \simeq 88\arcdeg$ ring will provide us with a powerful probe of intrinsic anisotropy, foregrounds, and point source behavior for our fields.

\subsubsection{CMB Polarization: Power Spectrum Detection Observing Mode}

The ring observing strategy described above will provide a robust detection of the polarization variance, or equivalently, of the \emph{amplitude} of the polarization power spectrum. In order to detect the \emph{shape} of the polarization power spectrum we are led to consider an observing strategy that allows for higher resolution in multipole-space. For the power-spectroscopy mode we note that the precision to which a given band of multipoles (or
band power $\hat{C_{\ell}})$, of width $L$, can be measured is given by \citep{knox95, hobson96, tegmark97},

$$\Delta \hat{C_{\ell}} \simeq \sqrt{\frac{2}{(2\ell +1)L f_{sky}}}
\Big[C_\ell + \frac{f_{sky}    w_P^{-1}  } {B^2_\ell}\Big],$$
where $f_{sky} =  N_{pix} \theta_{b}^2 / 4\pi$, and $\theta_{b}$ is the instrumental Gaussian beam's dispersion, which is related to its full width at half maximum by $\theta_{fwhm} = \sqrt{8\ln 2} \theta_b$. The optimum sky fraction for an experiment seeking to detect band power $\hat{C_{\ell}}$ at $\ell$ is given by \citep{tegmark97},

\beq f^{opt}_{sky} \simeq \frac{N_{pix} B^2_{\ell}
 C_{\ell}}{4\pi\sigma^2},
 \eeq
 where $\sigma$ is the effective noise per pixel in the final map. We see that the optimum sky fraction depends on the multipole. The band power that is measured with minimum total variance (cosmic + noise) occurs at the multipole where the cosmic variance equals the noise variance. For an experiment with fixed detector noise and observing time, the noise variance depends only on the beam size.  Since the cosmic variance is proportional to the power spectrum itself, this optimization suggests that the signal-to-noise ratio per beam-sized pixel should be approximately unity \citep{tegmark97}. The multipole corresponding to the angular scale of the beam-size is $\ell = 1/\theta_b$, which for the Polatron's $\theta_{fwhm}=2\arcmin$ beam implies that maximum sensitivity to the power spectrum is at $\ell \simeq 4000$. Unfortunately, the polarization power spectrum at $\ell = 4000$ is negligibly small compared to the maximum power which occurs at $\ell < 1000$. This leads us to consider enlarging the beam-size, either by co-adding multiple $2\arcmin$ pixels, or changing the receiver's feed horn. This latter alternative has the additional advantage of a more aggressive edge taper which further reduces side-lobe spillover from the ground. The optimal beam size for
power-spectrum detection is $6\arcmin < \theta_{fwhm} < 20\arcmin$, which is three to ten times larger than the beam size for the initial detection mode described above. The implementation of a larger beam will be implemented after an
initial polarization detection is made.

\section{Laboratory Characterization}
The optical properties of the Polatron receiver have been characterized in the laboratory.  An Infrared Labs HDL-8 liquid He dewar outfitted with a Chase Laboratories $^{3}$He sorption cooler provides a 0.300 K operating point for the bolometers.  Measurements of the spectral pass-band, optical efficiency, polarization efficiency, sensitivity, and detector properties for the entire system, including the wave plate, are presented here.

\subsection{Spectral Bands}
Spectral bands $\eta(\nu)$ were measured independently for the two polarization channels using a Fourier Transform Spectrometer (Figure 6).  The bands are identical within the 2 GHz spectral resolution of the FTS.  The center frequency $\nu_{0}$ of the band when observing a source of intensity $I_{\nu}$ is
\begin{equation}
\nu_{0} = \frac{\int\,d\nu\, \nu \, \eta(\nu)}{\int\,d\nu \, \eta(\nu)}.
\end{equation}
The measured band centers and widths are $\nu_{0} = 98 \pm 2$ GHz and $\Delta\nu = 17 \pm 2$ GHz full width at half maximum (FWHM) for both channels. 

We searched for spectral leaks at frequencies above our band with brass thick grill filters.  Thick grill filters \citep{timusk81} are metal plates with drilled holes that act as waveguide chokes at wavelengths above 1.7 times the hole diameter.  High-frequency radiation passes through, though with reduced efficiency due to the effective area of the holes.  A chopped 300 K / 77 K blackbody load was placed in front of the receiver and the signal was demodulated at the chop frequency.  A thick grill filter with calculated low-frequency cutoff at 156 GHz was used.  A measured upper limit of 1.0\% was placed on out-of-band power compared to in-band power for each channel. 

These measurements can be translated into an upper limit on the out-of-band contribution from dust to the brightness temperature that would be measured by the radiometer.  Since the detailed spectrum of any leak would be unknown, a worst-case scenario is assumed.  We represent high-latitude dust emission by a Rayleigh-Jeans spectrum with anisotropy temperature 70 $\mu$K (following \citet{ruhl93}) and 10\% polarization, an approximation which overestimates the expected power at all frequencies.  If the emission from the 77 K and 300 K loads is also assumed to be Rayleigh-Jeans at all frequencies, we can use the thick-grill measurement to set an upper limit of T = 0.07 $\mu$K to the out-of-band contribution from dust.

\subsection{Optical Efficiency}
The spectral band-averaged optical (photon) efficiency $\eta_{opt}$ of the receiver is measured by filling the beam with 300 K and 77 K blackbody loads, measuring the power on each detector for each load, and comparing the measured power difference to the expected power difference,
\begin{equation}
\Delta P_{opt} \approx \Delta\nu \frac{\eta_{opt}}{2}(A \Omega)  \left[B_{\nu_{0}}(300 K) - B_{\nu_{0}}(77 K)\right].
\end{equation}
The factor of $1/2$ accounts for the split in polarization, $A \Omega = \lambda^{2}$ is the throughput of the system, and $B_{\nu}(T)$ is the Planck blackbody spectrum.  In this way, we measure band-averaged optical efficiencies of $20.0\% \pm 2.5\%$ for both channels.  Nearly all of the loss in our optical train takes place in the cold filtering and at the detector itself.  The wave plate is anti-reflection coated to minimize loss, achieving a measured insertion loss of $< 1\%$.  The feed horn, OMT, and associated waveguide components were separately measured warm on a high-frequency network analyzer and found to generate $<1\%$ return loss across the spectral band.  

\subsection{Polarization Efficiency}
To measure the polarization efficiency, we constructed a polarized source by placing a 10 cm diameter circular aperture grid of 0.05 mm diameter gold-plated tungsten wire with 0.19 mm grid spacing in front of a chopped 300 K / 77 K blackbody load.  A piece of absorbing material was placed in front of the grid to limit the aperture to $\sim 5$ cm circular diameter.  The source was placed on the optic axis, approximately 10 cm from the entrance of the receiver.  Locking-in to the chopped signal, the wave plate was rotated in 128 steps through $360^{\circ}$ (See Figure 7).  Since $>99.5\%$ polarization is produced by the wire grid \citep{lesurf90}, the degree to which the lock-in signal does not fall to zero is the measured cross-polarization $\chi = 1 - \eta_{pol}$.  In this way we measure a laboratory polarization efficiency of $\eta_{pol} = 97.8\% \pm 0.7\%$, in good agreement with our calculation (see Appendix A.1).

The polarization efficiency of the receiver optics was measured in a similar manner, with the wave plate removed and the wire grid source rotated through $360^{\circ}$.  The measured polarization efficiency was $\eta_{pol} = 99.5\% \pm 0.7\%$, confirming that the dominant source of cross-polarization in our receiver is the wave plate.  The OMT and associated waveguide components were separately measured warm on a high-frequency network analyzer and found to generate $<0.03\%$ cross-polarization across the spectral band.  

Bolometer responsivity depends on the amount of optical loading.  For a large wire-grid source, rotation of the grid through $360^{\circ}$ could modulate the amount of power on the bolometer, since the bolometers can detect their 4 K optics in reflection.  This would under-estimate the polarization efficiency since the bolometers would be more responsive during the measurement of highest cross-polarization.  By limiting the size of the source and measuring the DC-level of the bolometers (in order to track the responsivity), we could be sure that the bolometer responsivity remained the same throughout the measurement. 

\subsection{Detector Properties}
The Polatron bolometers have thermal and electrical properties chosen to provide high sensitivity under the optical loading conditions and operating temperatures specific to the experiment.  They were previously flown as science detectors aboard the Boomerang payload in 1998.  The bolometers have a measured response time $\tau = 25$ ms, which puts an upper limit on the allowable frequency of modulation of the signal at $f \sim (2\pi\tau)^{-1} \sim$ 6 Hz.  The measured thermal conductivities G are $\sim 80$ pW/K, and the peak electrical responsivities S are $\sim 3 \times 10^{8}$ V/W \citep{crill00} under loading conditions similar to those expected at the telescope.

\subsection{Sensitivity}
The differential noise performance of the test receiver is adequate for optical testing purposes, but not stable enough for observation at the telescope.  For this receiver, the bolometers are DC-biased, and no efforts are made to mitigate radio-frequency pickup or microphonic pickup.  Furthermore, no trim circuit is available to balance the responsivities of the detectors, meaning that we are more susceptible to fluctuations in the bolometer stage temperature and other common mode signal.  

The noise spectrum was measured looking into a 77 K blackbody load, which is a fair approximation to the $\sim 60$ K loading temperature expected at the telescope.  The low-frequency signal is dominated by 1/f noise with a knee at $\sim$ 1 Hz.  The measured bolometer NEP at $f_{wp} =$ 0.6 Hz is $\sim 30 \times 10^{-17}$ W Hz$^{-1/2}$.  As calculated in $\S2.5$, and under similar observing conditions, this would amount to an NET$_{cmb} \sim 8$ mK s$^{1/2}$.  If we were observing with this test receiver, we would choose a higher waveplate rotation frequency, out of the 1/f noise but beneath the low-pass cutoff due to the bolometer response time.  Such a minimum in the noise spectrum exists at 4 Hz.  There we measure a bolometer NEP of $\sim 10 \times 10^{-17}$ W Hz$^{-1/2}$, which would amount to an NET$_{cmb} \sim 2.5$ mK s$^{1/2}$.  

\section{Conclusion}
We have described the Polatron experiment designed to measure the polarization angular power spectrum of the cosmic microwave background.  The Polatron focal plane is currently undergoing cryogenic integration and laboratory testing, after which the instrument will be installed at the OVRO 5.5m telescope.

\acknowledgments
The authors wish to thank the Caltech Physics Machine Shop for fabrication of parts and Kathy Deniston, Ken Ganga, Eric Hivon, and Marcus Runyan for many useful discussions.  This work was partly supported by NASA Innovative Research Grant NAG5-3465 and NASA NAG5-6573 for US involvement in Planck.  The bolometers were fabricated at the JPL Center for Space Microelectronics Technology.  SEC and BJP acknowledge a Stanford Terman Fellowship.  BJP was supported by a National Science Foundation Graduate Fellowship.  The construction of the Polatron is funded by a Caltech/JPL President's Fund Grant PF-414 and NSF Grant AST-9900868.  

\appendix
\section{Half-Wave Plates}
In this Appendix we review the optical properties of half-wave plates and calculate the expected polarization efficiency of our plate.  

The wave plate is a cylinder of birefringent, $x$-cut crystalline quartz.  The plate generates a $\pi$ phase retardation between electric field vectors of frequency $\nu_{0} \sim$ 96 GHz (wavelength $\lambda \sim 3$ mm) incident on the fast and slow refraction axes of the quartz.  The measured difference in index of refraction along the two axes at these wavelengths is $\Delta n = 0.048$, where $\bar{n} \sim 2.08$.  The differential phase shift $\Delta\phi$ generated between orthogonally polarized rays aligned with the fast and slow axes of the plate is 
\begin{equation}
\Delta\phi = \frac{2\pi\nu}{c}\Delta n \times l,
\end{equation}
where $l$ refers to the physical path length the rays travel through the plate.  For plane waves on the optic axis, $l$ is the thickness of the plate $t = 32.6$ mm, and a phase retardation of $\pi$ is achieved for $\nu = \nu_{0}$, the center of our band.  

Consider a linearly polarized plane wave traveling in the $\hat{z}$ direction in a coordinate system $({\bf\hat{x}},{\bf\hat{y}})$ fixed to our polarimeter with electric field
\begin{equation}
\overrightarrow{E}(x,y,z,t) = [E_{0x}{\bf\hat{x}} + E_{0y}{\bf\hat{y}}]\sin(\phi_{z,t}),
\end{equation}
where $\phi_{z,t} = 2\pi(\frac{z}{\lambda} - \nu t)$ is the kinematic component of the phase.

The intensity of such a beam is $I = E_{0x}^{2} + E_{0y}^{2}$, and the Stokes parameters measured by an ideal polarimeter aligned with this coordinate system are $Q_{0} = E_{0x}^{2} - E_{0y}^2$ and $U_{0} = 2 E_{0x} E_{0y}$.  

The wave is incident on a half-wave plate with polarization angle $\theta = \tan^{-1}(E_{0y}/E_{0x})$ with respect to the fast axis.  $\theta$ is restricted to the range $[0,\pi]$.  A second coordinate system, $({\bf\hat{i}}, {\bf\hat{j}})$, is fixed to the plate's fast and slow axes, respectively.  In the plate coordinates, the electric field vector of the incoming wave is 
\begin{equation}
\overrightarrow{E}(i,j,z,t) = \left\{ [E_{0x}\cos(\theta) + E_{0y}\sin(\theta)]{\bf\hat{i}} + [E_{0y}\cos(\theta) - E_{0x}\sin(\theta)]{\bf\hat{j}} \right\} \sin(\phi_{z,t})
\end{equation}
Upon passage through the plate, the phase of the ${\bf\hat{j}}$ component is delayed by $\Delta\phi = \frac{2\pi\nu}{c}\Delta n \times t$ with respect to the ${\bf\hat{i}}$ component.  For $\nu = \nu_{0}$, $\Delta\phi = \pi$ and so the ${\bf\hat{j}}$ term simply flips sign.  In other words, the polarization vector has been reflected about the fast axis, which is equivalent to a rotation by $\Delta\theta = -2\theta$.  Converting back to the polarimeter coordinate system,
\begin{equation}
\overrightarrow{E'}(x,y,z,t) = \left\{ [E_{0x}\cos(2\theta) + E_{0y}\sin(2\theta)]{\bf\hat{x}} + [E_{0y}\cos(2\theta) - E_{0x}\sin(2\theta)]{\bf\hat{y}} \right\} \sin(\phi_{z,t}).
\end{equation}
The Stokes parameters are rotated by $4\theta$:
\begin{eqnarray}
Q' & = & {[E_{0x}\cos(2\theta) + E_{0y}\sin(2\theta)]}^2 - {[E_{0y}\cos(2\theta) - E_{0x}\sin(2\theta)]}^2 \\
   & = & Q_{0}\cos(4\theta) + U_{0}\sin(4\theta),
\end{eqnarray}
and
\begin{equation}
U' = U_{0}\cos(4\theta) - Q_{0}\sin(4\theta).
\end{equation}

\subsection{Wave plate cross-polarization for $\nu \neq \nu_{0}$}
Consider a polarized beam of intensity $I_{0}$ incident on the wave plate with polarization vector ${\bf\hat{y}}$ aligned at a $45^{\circ}$ angle to the fast $({\bf\hat{i}})$ and slow $({\bf\hat{j}})$ axes of the plate.  For frequency $\nu_{0}$ the polarization will be rotated by $90^{\circ}$, into the ${\bf\hat{x}}$ direction, and a polarization detector aligned with ${\bf\hat{y}}$ will measure no signal.  For frequencies $\nu \neq \nu_{0}$ the wave plate is less efficient, and some cross-polar signal will be detected.  The electric field incident on the plate is
\begin{equation}
\overrightarrow{E}(i,j,z,t) = \frac{E_{0}}{\sqrt{2}}\sin(\phi_{z,t})\left[{\bf\hat{i}} + {\bf\hat{j}}\,\right].
\end{equation}
The phase retardation for the polarization component aligned with the slow axis is $\Delta\phi = \frac{2\pi\nu}{c} \Delta n \times t = \pi + \delta\phi$, where $\delta\phi = \pi((\nu/\nu_{0}) - 1)$.  Upon passage through the plate,
\begin{eqnarray}
\overrightarrow{E}(i,j,z,t) & = & \frac{E_{0}}{\sqrt{2}}\left[\sin(\phi_{z,t}){\bf\hat{i}} +
                                  \sin(\phi_{z,t}+\pi+\delta\phi)\,{\bf\hat{j}}\,\right], \\
                            & = & \frac{E_{0}}{\sqrt{2}}\left[\sin(\phi_{z,t}){\bf\hat{i}} -
                                  \sin(\phi_{z,t} + \delta\phi){\bf\hat{j}}\,\right].
\end{eqnarray}
The projection of this vector along the detector (${\bf\hat{y}}$) direction is 
\begin{eqnarray}
\overrightarrow{E} \cdot {\bf\hat{y}} & = & \frac{E_{0}}{2}\left[\sin(\phi_{z,t}) - \sin(\phi_{z,t} + \delta\phi)\right] \\
& = & \frac{E_{0}}{2}\left[\sin(\phi_{z,t}) - \sin(\phi_{z,t})\cos(\delta\phi) - \cos(\phi_{z,t})\sin(\delta\phi)\right].
\end{eqnarray}
The detector is sensitive to the time average of the square of the electric field,
\begin{equation}
\langle |\overrightarrow{E} \cdot {\bf\hat{y}}|^{2} \rangle = I_{0} \times \frac{1}{2}\left[1 - \cos(\delta\phi)\right].
\end{equation}
The total fractional cross-polarization seen by a detector with spectral response $\eta(\nu)$ aligned along the ${\bf\hat{y}}$ direction is then
\begin{equation}
\label{chi}
\chi = \frac{\int_{0}^{\infty}d\nu \,\eta(\nu) I_{0}(\nu) \times \frac{1}{2} [1-\cos(\delta\phi)]}
{\int_{0}^{\infty} d\nu \,\eta(\nu)I_{0}(\nu)}.
\end{equation}
With the measured spectral response of the Polatron receiver, Eqn.(\ref{chi}) predicts a cross-polarization $\chi = 0.9\% \pm 0.1\%$ ($\eta_{pol} = 1 - \chi = 99.1\% \pm 0.1\%$), where the error is dominated by the accuracy of the measurement of the spectral band near the band edges.  The same calculation with a model Gaussian spectral band with band center and width equal to the measured Polatron values predicts $\chi = 0.8\%$ ($\eta_{pol} = 99.2\%$).

\subsection{Wave plate cross-polarization due to converging beam}
Placement of the wave plate in a converging beam generates cross-polarization since different parts of the beam travel through different thicknesses of quartz, and experience different phase shifts.  Modeling this effect can be quite involved, since the wave plate is in the near field of a single-mode gaussian beam with a changing phase profile.  Since the expected cross-polarization is fairly low, a simple calculation can help us to understand the basic performance.

Consider a polarized beam of intensity $I_{0}$ incident on the wave plate at angle $\theta$ with respect to normal, and with polarization vector ${\bf\hat{y}}$ aligned at a $45^{\circ}$ angle to the fast $({\bf\hat{i}})$ and slow $({\bf\hat{j}})$ axes of the plate.  The phase shift upon passage through the plate is $\Delta\phi = \frac{2\pi\nu}{c} \Delta n \times t \csc(\theta) = \pi + \delta\phi$, where $\delta\phi = \pi(\csc(\theta) - 1)$.  The response of a polarization detector aligned with ${\bf\hat{y}}$ is as shown in equation (A13).  

The angular response $g(\theta)$ of the feed horn at a distance $z$ from the horn phase center (focus) can be estimated as in \citet{wylde84}:
\begin{equation}
g(\theta) \propto e^{- 2 \theta^{2} (z / w(z))^{2}},
\end{equation}
where
\begin{equation}
w(z) = w_{0} \times \left[ 1 + \left(\frac{\lambda z}{\pi w_{0}^{2}}\right)^{2}\right]^{1/2},
\end{equation}
$w_{0} = 0.64\,a$, and $a$ is the aperture radius of the feed horn.

The total fractional cross-polarization seen by the detector is then
\begin{equation}
\chi = \frac { \int_{ -\frac{\pi}{2}}^{\frac{\pi}{2}} d\theta \, 2\pi\theta\, g(\theta) \times \frac{1}{2} [1-\cos(\delta\phi)]}
{ \int_{-\frac{\pi}{2}}^{\frac{\pi}{2}} d\theta \, 2\pi\theta \,g(\theta)}.
\end{equation}

This predicts $\chi \sim 2\%$ ($\eta_{pol} = 98\%$) for the Polatron receiver.  In combination with the calculation of cross-polarization due to the Polatron spectral bandwidth, a total polarization efficiency for the receiver of $\eta_{pol} \sim 97\%$ is predicted.  The measured efficiency is 97.8\% (\S3.3).

\newpage
\textbf{Table Captions}
\begin{enumerate}
\item Polatron Instrument Specifications
\item Polatron Anticipated Sensitivity
\end{enumerate}

\newpage
\begin{deluxetable}{ll}
\tablewidth{0pt}
\tablecaption{Instrument Specifications}
\tabletypesize{\small}
\tablenum{1}
\label{specs}
\tablecolumns{2}
\startdata
\tableline
\\
Telescope: & 5.5 m Cassegrain Focus \\
\\
Polarization Analysis: & Rotating Wave Plate and \\ 
                       & Orthomode Transducer    \\
\\
Polarization Efficiency: & 97.8\%  \\ 
\\
Spectral Band: & 88 -- 106 GHz \\ 
\\
Beam Size: & 2.5 arcminutes FWHM  \\ 
\\
Detectors: & Silicon-Nitride Micromesh \\ & Bolometers \\ 
\\
Sensitivity: & $\sim 700~\mu$K s$^{1/2}$ to both \textit{Q} and \textit{U} \\ 
\\
\enddata
\end{deluxetable}

\newpage

\begin{deluxetable}{cccccccc}
\tablewidth{0pt}
\tablecaption{Anticipated Sensitivity}
\label{sens}
\tablenum{2}
\tablecolumns{8}
 
\startdata

 \tableline

 \multicolumn{1}{c}{\textit{ppwv}} &
 \multicolumn{5}{c}{NEP $10^{-17}$ W Hz$^{-1/2}$} &
 \multicolumn{1}{c}{NEFD} &
 \multicolumn{1}{c}{NET} \\ 

 \multicolumn{1}{c}{(mm)} & 
 \multicolumn{1}{c}{Photon} & 
 \multicolumn{1}{c}{Detector} & 
 \multicolumn{1}{c}{Amp.} &
 \multicolumn{1}{c}{Atmos.} & 
 \multicolumn{1}{c}{Total} & 
 \multicolumn{1}{c}{mJy s$^{1/2}$} &
 \multicolumn{1}{c} {$\mu$K s$^{1/2}$} \\ 

 \tableline

 3.0 & 2.5 & 2.2 & 1.0 & 1.0 & 3.6 & 85 & 610 \\
 6.0 & 2.7 & 2.2 & 1.0 & 1.2 & 3.9 & 99 & 710 \\
 8.0 & 3.0 & 2.2 & 1.0 & 1.4 & 4.1 & 101 & 740 \\ 

 \tableline

\enddata

\end{deluxetable}

\newpage

\begin{figure} 
\epsscale{0.5}
\plotone{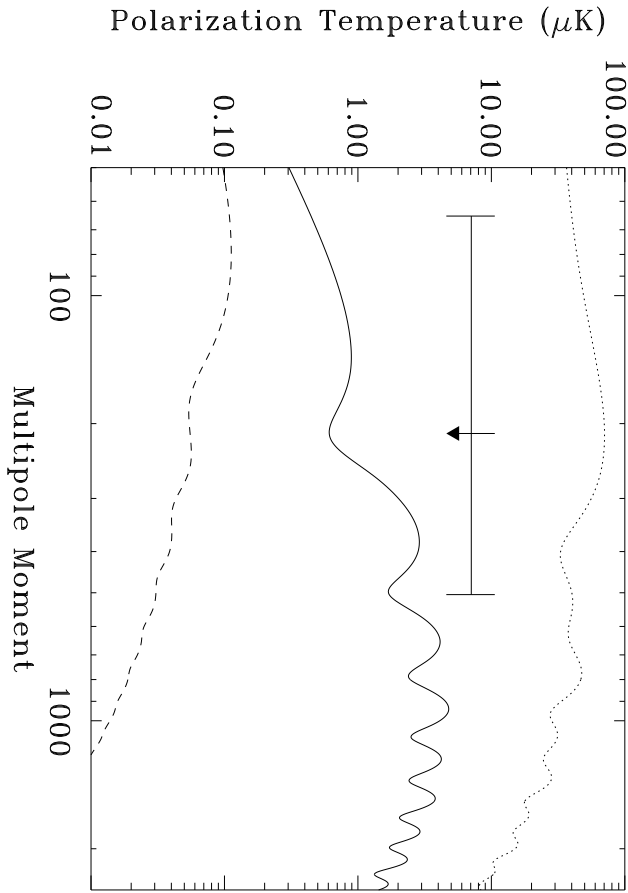}

\caption{Model Angular Power Spectra.  Model angular power spectra generated by CMBFAST (see www.physics.nyu.edu/matiasz/CMBFAST/cmbfast.html) are plotted for an $\Omega = 1, \Omega_{B} = 0.05, \Omega_{CDM} = 0.27, H = 81 Km/s/MPc, n_{S} = 0.97, N_{T} = 1 - n_{S}$, and $T/S = 7(1 - n_{S}) = 0.21$ universe.  Solid line is G-type polarization, dashed line is C-type polarization, and dotted line is best fit temperature power spectrum to Boomerang data.  Upper limit is from \citet{hedman00}, assuming $C_{l}^{C} = 0$, as described in text.}
\end{figure}

\begin{figure} 
\plotone{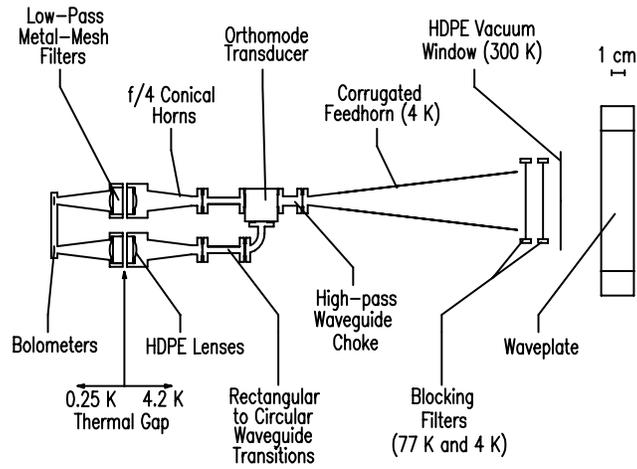} 
\caption{Polatron Focal Plane Schematic.  A single-mode entrance feed is coupled to an orthomode transducer, which separates the two senses of linear polarization.  Each OMT output is spectrally filtered and then terminated in a ``spider-web'' bolometer.}
\end{figure}

\begin{figure} 
\plotone{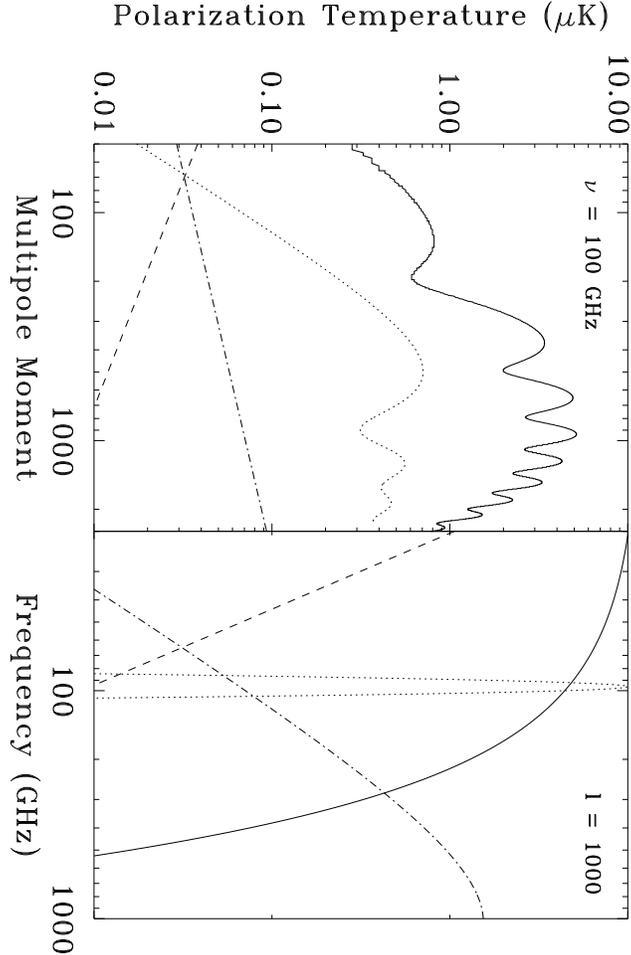} 
\caption{Polarization Foreground Spectra.  Left panel:  predicted polarization flat band power at 100 GHz as a function of multipole moment $l$ for model CMB G-type fluctuations (solid line), galactic synchrotron (dashed line), and galactic dust (dash-dotted line).  The dotted line indicates a generic Polatron window function in arbitrary units (see \S2.9 for details.)  Right panel:  predicted polarization flat band power at $l$ = 1000 as a function of spectral frequency, same linetype key, with dotted line indicating the chosen Polatron spectral passband in arbitrary units.  Temperatures are stated in Rayleigh-Jeans units.  CMB G-type angular spectrum was generated by CMBFAST for a CDM ($\Omega = 1, \Omega_{\Lambda} = 0.6, \Omega_{B} = 0.05, H = 65$) universe.}
\end{figure} 

\begin{figure} 
\plotone{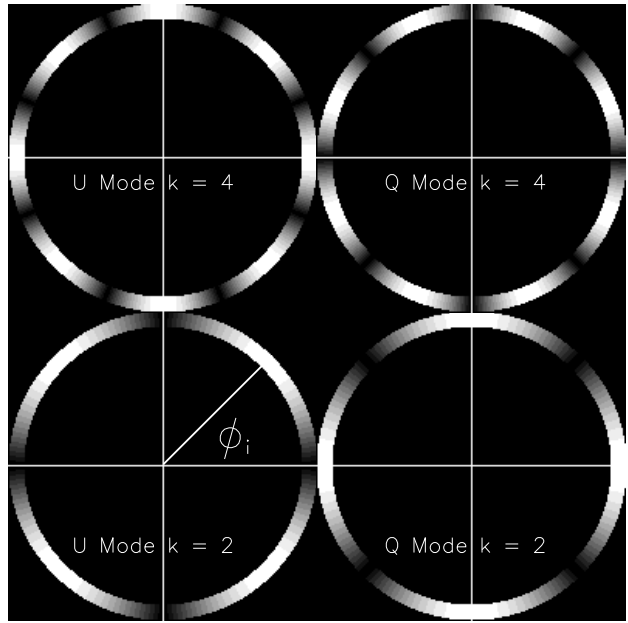} 
\caption{Polatron Observing Strategy. This figure illustrates the Stokes parameter weight functions, as described in the text. The functions are shown real space, for both Q and U, for four example modes. The rings are all centered on the North Celestial Pole (NCP), at a declination of $\delta =\theta$. The azimuthal Right Ascension coordinates of the map pixels are indexed by $\phi_i$.}
\end{figure}

\begin{figure} 
\plotone{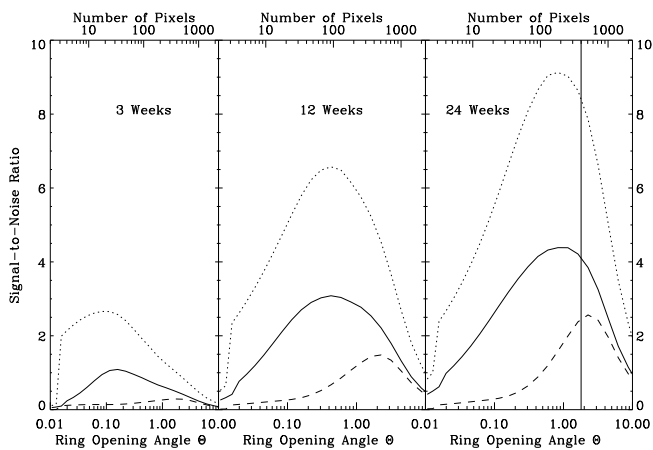} 
\caption{Determination of Ring Opening Angle.  Plotted are the Polatron integrated signal to noise ratio for three different integration times over a range of ring opening angles.  The dotted curves show the performance of an 
ideal instrument with no 1/f noise that measures both Q and U simultaneously.  The dashed curves show the performance of an  instrument with 1/f noise that measures both Q and U simultaneously, but does not difference pairs of Stokes parameters separated by 6h in position angle around the ring.  The dashed curves show the performance of an 
instrument with 1/f noise that measures both Q and U simultaneously, but does difference pairs of Stokes parameters separated by 6h in position angle around the ring.}
\end{figure} 

\begin{figure} 
\plotone{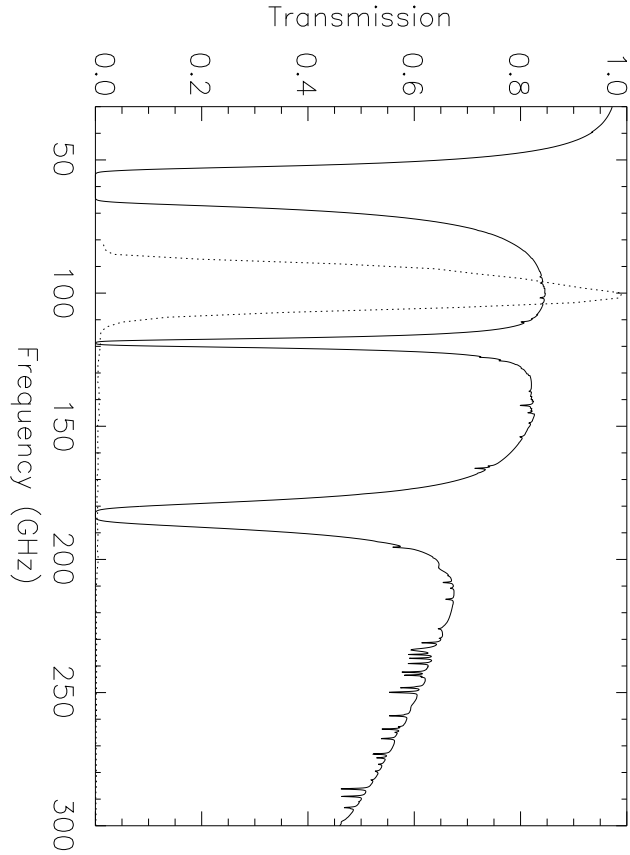} 
\caption{Atmospheric Spectrum and Polatron Passband.  Solid line:  a model atmospheric transmission spectrum for the Owens Valley assuming 6 mm of precipitable water vapor.  Dotted line:  the Polatron passband, in units relative to peak transmission, as measured on a Fourier Transform Spectrometer (FTS).  The frequency resolution of the FTS is 2 GHz.  The spectral bands for the two orthogonal polarization channels are identical within this resolution.}
\end{figure}

\begin{figure} 
\plotone{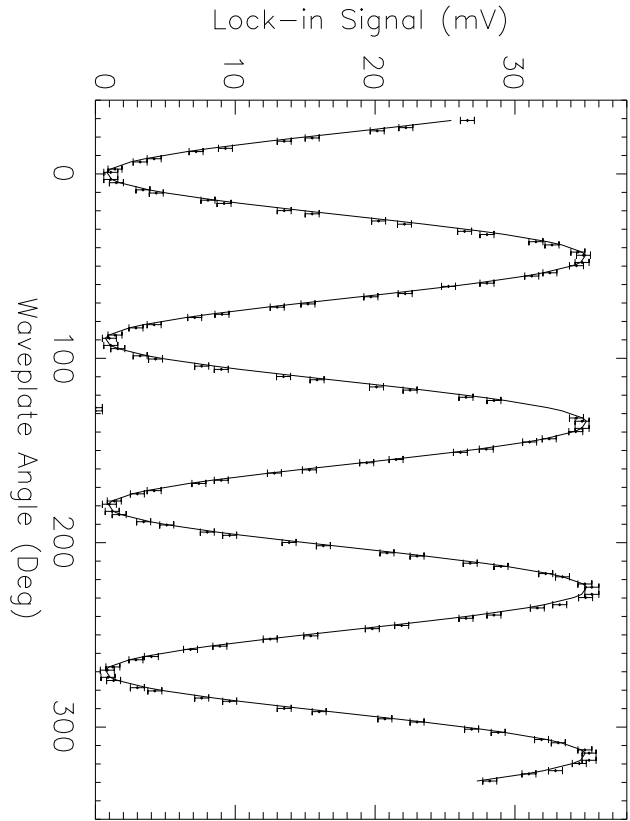} 
\caption{Polarization Efficiency Measurement.  Depicted is the result of the polarization efficiency measurement described in \S3.3.  Error bars are 2-$\sigma$.  Best fit function is $0.8 + 34.4\sin(4f) - 0.2\sin(2f - 5.6^{\circ}) + 0.1\sin(f)$ mV; the measured polarization efficiency is $(97.8 \pm 0.7)\%$.}
\end{figure}


\begin{thebibliography}{}

\bibitem[Bhatia et al.(2001)]{bhatia01}Bhatia, R.S., et al., in Proc. of the 11th International
Cryocooler Conf., ed. R.G. Ross, Jr. (New York: Plenum Press), 577
\bibitem[Bhatia et al.(1999)]{bhatia99}Bhatia, R.S., et al. 1999, Cryogenics, 39, 701
\bibitem[Bouchet et al.(1999)]{bouchet98}Bouchet, F.R., et al. 1999, New Astronomy, 4, 443
\bibitem[Chandrasekar(1960)]{chandra60}Chandrasekar S. 1960, Radiative Transfer (New York: Dover)
\bibitem[Church(1995)]{church95}Church, S.E. 1995, \mnras, 272, 551
\bibitem[Church \textit{et al.}(1996)]{Church96}Church, S.E., Philhour, B.J., Lange, A.E., Ade, P.A.R., Maffei, B., Nartallo-Garcia, R., \& Dragovan, M. 1996, in Conf. Proc. 30th ESLAB Symposium (Noordwijk:  ESTEC) 
\bibitem[Crill(2000)]{crill00}Crill, B.P. 2000, Ph.D. Thesis (California Institute of Technology)
\bibitem[Crittenden, Coulson, \& Turok(1995)]{crittenden95}Crittenden, R.G., Coulson, D., \& Turok, N.G. 1995, \prd, 52, 5402
\bibitem[Danese \& Partridge(1989)]{danese89}Danese, L., \& Partridge, R.B. 1989, \apj, 342, 604
\bibitem[de Bernardis et al.(1999)]{deBernardis99}de Bernardis, P., et al. 1999, New Astronomy Reviews, 43, 289
\bibitem[Delabrouille, Gorski, \& Hivon(1998)]{delabrouille98}Delabrouille, J., Gorski, K.M., \& Hivon, E. 1998, \mnras, 298, 445
\bibitem[Draine \& Lazarian(1998)]{draine98}Draine, B., \& Lazarian, A. 1998, \apj, 494, L19
\bibitem[Hecht \& Zajac(1982)]{hecht82}Hecht, E., \& Zajac, A. 1982, Optics (Reading:  Addison-Wesley)
\bibitem[Hedman et al.(2001)]{hedman00}Hedman, M. M., Barkats, D., Gundersen, J. O., Staggs, S. T., \& Winstein, B. 2001, \apj, 548, L111
\bibitem[Herbig et al.(1995)]{herbig95}Herbig, T., Lawrence, C. R., Readhead, A. C. S., \& Gulkis, S. 1995, \apjl, 449, L5
\bibitem[Hildebrand \& Dragovan(1995)]{hildebrand95}Hildebrand, R., \& Dragovan, M. 1995, \apj, 450, 663
\bibitem[Hobson, M.P. \& Magueijo, J.(1996)]{hobson96}Hobson, M.P. \& Magueijo, J. 1996, \mnras, 283, 1133
\bibitem[Holzapfel et al.(1997)]{holzapfel97}Holzapfel, W. L., Wilbanks, T. M., Ade, P. A. R., Church, S. E., Fischer, M. L., Mauskopf, P. D., Osgood, D. E., \& Lange, A. E. 1997, \apj, 479, 17
\bibitem[Hu \& White(1997)]{hu97}Hu, W., \& White, M. 1997, New Astronomy, 2, 323
\bibitem[Glenn(1997)]{glenn97}Glenn, J. 1997, Ph.D. Thesis (University of Arizona)
\bibitem[Jaffe, Kamionkowski, \& Wang(2000)]{jaffe00} Jaffe, A.\ H., Kamionkowski, M., \& Wang, L.\ 2000, \prd, 61, 11462d 
\bibitem[Kamionkowski, Kosowsky, \& Stebbins(1997a)]{kamionkowski97}Kamionkowski, M., Kosowsky, A., \& Stebbins, A.\ 1997, \prd, 55, 7368 
\bibitem[Kamionkowski, Kosowsky, \& Stebbins(1997b)]{kamionkowski97b}Kamionkowski, M., Kosowsky, A., \& Stebbins, A.\ 1997, \prl, 78, 2058 
\bibitem[Keating et al.(1998)]{keating98} Keating, B., Timbie, P., Polnarev, 
A., \& Steinberger, J.\ 1998, \apj, 495, 580 
\bibitem[Knox, L.(1995)]{knox95}Knox, L. 1995, \prd, 52, 4307
\bibitem[Lange et al.(2000)]{lange00}Lange, A.E., et al. 2001, \prd, in press (astro-ph/0005004)
\bibitem[Lawrence, Herbig, \& Readhead(1994)]{lawrence94}Lawrence, C.L., Herbig, T., \& Readhead, A.C.S. 1994, Proc. IEEE, 82, 763
\bibitem[Lee, Ade, \& Haynes(1996)]{Lee96}Lee, C., Ade, P.A.R., \& Haynes, C.V. 1996, in Conf. Proc. 30th ESLAB Symposium (Noordwijk:  ESTEC) 
\bibitem[Leitch et al.(2000)]{leitch00} Leitch, E.\ M., Readhead, A.\ C.\ S., Pearson, T.\ J., Myers, S.\ T., Gulkis, S., \& Lawrence, C.\ R.\ 2000, \apj, 532, 37 
\bibitem[Lesurf(1990)]{lesurf90}Lesurf, J.C.G. 1990, Millimetre-wave Optics, Devices and Systems (Bristol: Adam Hilger)
\bibitem[Mason, Myers, \& Readhead(2001)]{mason01}Mason, B.\ S., Myers, S.\ T., \& Readhead, A.\ C.\ S. 2001, \apj, in press (astro-ph/0101169)
\bibitem[Murray et al.(1992)]{murray92}Murray, A.G., et al. 1992, Infrared Physics, 33, 113
\bibitem[Myers, Readhead, \& Lawrence(1993)]{myers93}Myers, S.\ T., Readhead, A.\ C.\ S., \& Lawrence, C.\ R. 1993, \apj, 405
\bibitem[Myers et al.(1997)]{myers97} Myers, S.\ T., Baker, J.\ E., Readhead, A.\ C.\ S., Leitch, E.\ M., \& Herbig, T.\ 1997, \apj, 485, 1 
\bibitem[Netterfield et al.(1995)]{netterfield95} Netterfield, C.\ B., Jarosik, N., Page, L., Wilkinson, D., \& Wollack, E.\ 1995, \apjl, 445, L69 
\bibitem[Partridge, Nowakowski, \& Martin(1988)]{partridge88} Partridge, R.\ B., Nowakowski, J., \& Martin, H.\ M.\ 1988, \nat, 331, 146 
\bibitem[Prunet, Sethi, \& Bouchet(2000)]{prunet00} Prunet, S., Sethi, S.\ K., \& Bouchet, F.\ R.\ 2000, \mnras, 314, 348 \bibitem[Richards(1994)]{richards94}Richards, P.L., J. Appl. Phys. 1994, 76, 1
\bibitem[Readhead et al.(1989)]{readhead89}Readhead, A.\ C.\ S., Lawrence, C.\ R., Myers, S.\ T., Sargent, W.\ L.\ W., Hardebeck, H.\ E., \& Moffet, A.\ T. 1989, \apj, 346, 566
\bibitem[Ruhl(1993)]{ruhl93}Ruhl, J. 1993, Ph.D. Thesis (Princeton University)
\bibitem[Seljak \& Zaldarriaga(1997)]{seljak97} Seljak, U.\ \& Zaldarriaga, M.\ 1997, \prl, 78, 2054 
\bibitem[Staggs et al.(1999)]{staggs99}Staggs, S.\ T.\ et al.\ 1999, ASP Conf.\ Ser.\ 181: Microwave Foregrounds, 299 
\bibitem[Tegmark(1997)]{tegmark97}Tegmark, M. 1997, \prd, 56, 4514 
\bibitem[Tegmark \& de Oliveira-Costa(2000)]{tegmark00}Tegmark, M. \& de Oliveira-Costa, A. 2000, in preparation (astro-ph/0012120)
\bibitem[Timusk \& Richards(1981)]{timusk81}Timusk, T., \& Richards, P.L. 1981, \ao, 20, 1355
\bibitem[Wilbanks et al.(1990)]{wilbanks90}Wilbanks, T., Devlin, M., Lange, A.E., Sato, S., Beeman, J.W., \& Haller, E.E. 1990, IEEE Transactions on Nuclear Science, 37, 566
\bibitem[Wylde(1984)]{wylde84}Wylde, R.J. 1984, IEEE Proceedings, 131, 258
\bibitem[Zaldarriaga \& Seljak(1997)]{zaldarriaga97}Zaldarriaga, M. \& Seljak, U., 1997, \prd, 55, 1830
\bibitem[Zaldarriaga(1998)]{zaldarriaga98}Zaldarriaga, M. 1998, \apj, 503, 1

\end{thebibliography}
\end{document}